\documentclass[a4paper,11pt]{article}
\pdfoutput=1

\usepackage{jheppub}
\usepackage[utf8]{inputenc} 
\usepackage{amsfonts,amssymb,mathrsfs,amsmath,esint,bm}
\usepackage{pdflscape} 
\usepackage[capitalise]{cleveref}
\allowdisplaybreaks

\graphicspath{{./Figures/}}
\usepackage{latexsym}
\usepackage{graphicx}
\usepackage[dvipsnames]{xcolor}
\usepackage{booktabs}
\usepackage{physics}
\usepackage{tikz}
\usetikzlibrary{decorations.pathmorphing}\usetikzlibrary{positioning,calc}
\usetikzlibrary{math}
\usepackage{cleveref}
\usepackage{comment}
\usepackage[normalem]{ulem}
\newcommand{\stkout}[1]{\ifmmode\text{\sout{\ensuremath{#1}}}\else\sout{#1}\fi} 
\graphicspath{{./Figures/}}

\renewcommand{\d}{\mathrm{d}}
\newcommand{\parfrac}[2]{\left(\frac{#1}{#2}\right)}

\newcommand{\HI}{H_I}
\newcommand{\MP}{M_\textsc{p}}

\renewcommand{\epsilon}{\varepsilon}
\newcommand{\rh}{\textsc{rh}}

\renewcommand{\arg}{\text{arg}}
\newcommand{\GeV}{\,\text{GeV}}

\newcommand{\Nk}{n_k}
\newcommand{\nco}{n_{\chi,\rm co}}
\newcommand{\n}{n_{\chi}}
\newcommand{\aend}{a_\text{end}}
\newcommand{\tend}{t_\text{end}}
\newcommand{\kend}{k_\text{end}}
\newcommand{\Hend}{H_\text{end}}
\newcommand{\tfinal}{t_\text{final}}
\definecolor{poleCross}{rgb}{0,0.8,.13}
\definecolor{poleRH}{rgb}{0,0.84,1}
\definecolor{darkgreen}{rgb}{0,0.5,0}

\title{
Gravitational Production of Heavy Particles during and after Inflation
}
\author[a,b,c,1]{Davide~Racco,} 
\author[d,2]{Sarunas~Verner,} 
\author[d,3]{and~Wei~Xue}

\affiliation[a]{Institut f\"ur Theoretische Physik, ETH Z\"urich,\\ 
Wolfgang-Pauli-Str.\ 27, 8093 Z\"urich, Switzerland}

\affiliation[b]{Physik-Institut, Universit\"at Z\"urich,\\
Winterthurerstrasse 190, 8057 Z\"urich, Switzerland}

\affiliation[c]{Stanford Institute for Theoretical Physics, Stanford University,\\
382 Via Pueblo Mall, Stanford, CA 94305, USA}

\affiliation[d]{Institute for Fundamental Theory, Physics Department, University of Florida,\\ Gainesville, FL 32611, USA}

\emailAdd{dracco@phys.ethz.ch}
\emailAdd{verner.s@ufl.edu}
\emailAdd{weixue@ufl.edu}

\abstract{
We investigate the gravitational production of a scalar field $\chi$ with a mass exceeding the Hubble scale during inflation $m_\chi \gtrsim \HI$, employing both analytical and numerical approaches. 
We demonstrate that the steepest descent method effectively captures the epochs and yields of gravitational production in a compact and simple analytical framework.
These analytical results align with the numerical solutions of the field equation.
Our study covers three spacetime backgrounds: de Sitter, power-law inflation, and the Starobinsky inflation model.
Within these models, we identify two distinct phases of particle production: during and after inflation.
During inflation, we derive an accurate analytic expression for the particle production rate, accounting for a varying Hubble rate.
After inflation, the additional burst of particle production depends on the inflaton mass around its minimum.
When this mass is smaller than the Hubble scale during inflation, $\HI$, there is no significant extra production. 
However, if the inflaton mass is larger, post-inflation production becomes the dominant contribution. 
Furthermore, we explore the implications of gravitationally produced heavy fields for dark matter abundance, assuming their cosmological stability.
}

\begin{document}

\maketitle

\section{Introduction}
\label{sec:introduction}

Multiple features of the Universe that we observe on large scales point to the paradigm of primordial inflation as a framework that 
describes the initial conditions and early dynamics of the observable Universe 
(see e.g.~\cite{Lyth:1998xn, Senatore:2016aui, Baumann:2018muz} for reviews). 
Inflation marks an early stage of accelerated expansion in the universe. 
This rapid expansion serves as a generator of new particles, including those with minimal couplings to gravity 
\cite{Parker:1968mv, Parker:1969au, Parker:1971pt, Ford:2021syk}.

These particles arise through minimal and inevitable gravitational interactions,  
influencing diverse phenomena, such as dark matter, cosmological (iso)curvature perturbation, baryon asymmetry, and gravitational waves
(see \cite{Kolb:2023ydq} for a review). 
The existence of dark matter is supported by gravitational probes, but there is no evidence yet of other interactions, 
making the minimal gravitational production an attractive and motivated option.
This mechanism has been studied for dark matter consisting of a single particle that couples only through gravity, with different evolution histories depending on its
spin: 0~\cite{Chung:1998zb, Chung:1998ua, Kolb:1998ki, Ling:2021zlj, Garcia:2023qab},
$\tfrac 12$ \cite{Lyth:1996yj, Kuzmin:1998kk, Chung:2011ck},
1 \cite{Graham:2015rva, Ahmed:2020fhc, Kolb:2020fwh,Gorghetto:2022sue, Cembranos:2023qph,Ozsoy:2023gnl},
$\tfrac 32$ \cite{Hasegawa:2017hgd,Antoniadis:2021jtg, Kaneta:2023uwi,Casagrande:2023fjk},
2 \cite{Kolb:2023dzp}. 
Beyond dark matter, gravitational production inevitably yields a contribution to dark sector as a byproduct of inflation, as 
explored in \cite{Gross:2020zam, Redi:2020ffc,Krnjaic:2020znf, Arvanitaki:2021qlj, Redi:2022zkt,Redi:2022myr, East:2022rsi, Bastero-Gil:2023htv, Bastero-Gil:2023mxm}.
Furthermore, depending on the particle's mass, spin, and interactions, the (post-)inflationary production of particles 
may overclose the Universe (e.g.~in the case of light, weakly coupled scalar fields), produce non-Gaussianities, or excessive isocurvature fluctuations on the large scales probed by the cosmic microwave background (CMB) 
\cite{Chung:2004nh,Chung:2011xd,Chung:2013sla,Chung:2015pga,Markkanen:2017rvi, Herring:2019hbe,Garcia:2023awt}.

Gravitational particle creation occurs during inflation, and also after inflation.
The beginning of the thermal history of the Universe during the (p)reheating phase after inflation may provide further graviton-mediated particle production via scattering of massive inflaton modes or hot Standard Model (SM) particles  \cite{Greene:1998nh, Ema:2015dka, Garny:2015sjg, Ema:2016hlw, Tang:2017hvq, 
Garny:2017kha, Bernal:2018qlk, Ema:2018ucl, Garny:2018grs, Opferkuch:2019zbd, 
Chianese:2020yjo, Herring:2020cah, 
Mambrini:2021zpp, Bernal:2021kaj, Barman:2021ugy, Garani:2021zrr, Ahmed:2021fvt, Clery:2021bwz, Haque:2021mab, 
Haque:2022kez, Aoki:2022dzd, Clery:2022wib, Ahmed:2022tfm,  Garcia:2022vwm, Basso:2022tpd, Kaneta:2022gug, Barman:2022qgt,
Haque:2023yra, Figueroa:2024asq}.

Motivated by this rich phenomenology, and by the following theoretical inquiries, we investigate the gravitational production of a heavy scalar field $\chi$, with a mass $m_\chi$ larger than the inflationary Hubble rate $\HI$.

Phenomenologically, scalar masses are generally not protected by symmetries from large corrections, leading to the natural expectation that they are heavy, near the scale of new physics. 
When these particles are heavier than the Hubble rate during inflation, their spectrum of perturbations becomes blue-tilted (i.e.~enhanced on short scales), thus avoiding problematic isocurvature perturbations on the CMB scales. 
Moreover, the gravitational production of heavy scalars is typically suppressed, thereby preventing overproduction and 
rendering them promising candidates for dark matter.

In theoretical terms, several crucial questions regarding the production of a simple spin-0 particle in an inflationary background remain unanswered. 
These include: 
1) determining the particle production rate during inflation, accounting for the time variation of the Hubble rate; 
2) understanding the model dependence of heavy particle production; 
3) comprehending particle production during and after inflation within a unified framework.

Previous literature has primarily focused on particle production during inflation within de Sitter spacetime, or equivalently,  by assuming a constant Hubble rate, yielding a number density $\sim \exp ( - 2\pi {m_\chi}/ \HI )$
\cite{Markkanen:2016aes,Li:2019ves,Corba:2022ugu}.
However, the Hubble rate during inflation undergoes gradual changes. 
After 60 $e$-foldings, this variation can result in a significant modification in the predicted particle abundance, as the Hubble rate modifies exponentially the produced number density.
Additionally, the final abundance of heavy particles is predominantly influenced by the UV modes which are almost leaving the Hubble radius at the end of inflation, when the Hubble rate changes even more rapidly.
Therefore, this motivates us to explore the particle production with a slow-rolling Hubble scale during inflation.

After inflation, the coherent oscillations of the inflaton field around its minimum can trigger the production of $\chi$ through gravitational interactions. 
Previous studies \cite{Ema:2015dka,Ema:2016hlw, Ema:2018ucl} highlighted the possibility that gravitational particle production may not be exponentially suppressed when $m_\phi > m_\chi \gg \HI$, because of graviton-mediated pair annihilations of the inflaton $\phi$ into the scalar field $\chi$. 
Some analytic treatment was developed in \cite{Chung:2018ayg}.
A gap between the inflaton mass in its minimum, and the inflationary $\HI$, can be realized in the hilltop inflationary model. 
In our study, we focus on the complementary parameter space, where the inflaton mass is smaller than or comparable to the spectator mass, i.e.~$m_\phi \lesssim  m_\chi$ and  $ m_\chi  \gtrsim \HI$. 
Notably, this parameter space can naturally arise in various inflationary models. 

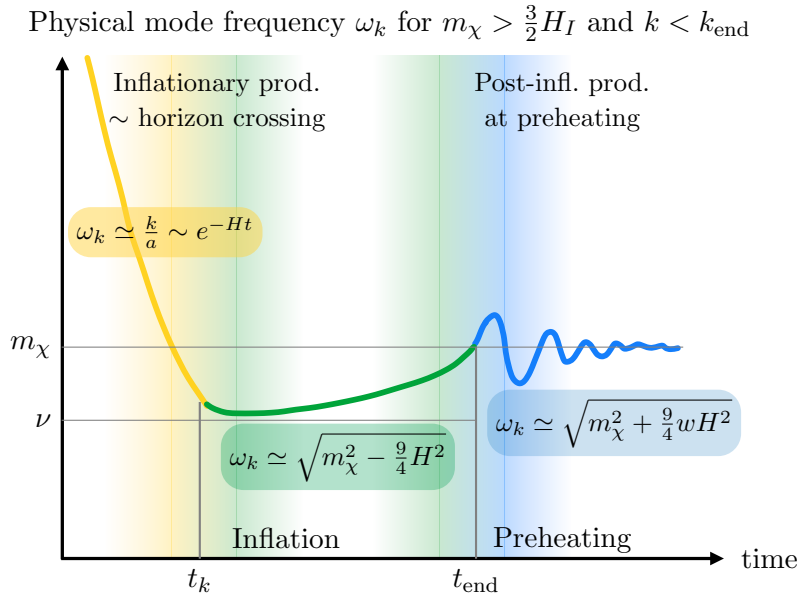
\begin{figure}\centering
\begin{tikzpicture}
\node [above right,inner sep=0] (image) at (0,0) {\includegraphics[width=.6\textwidth,trim=540 220 540 220,clip]{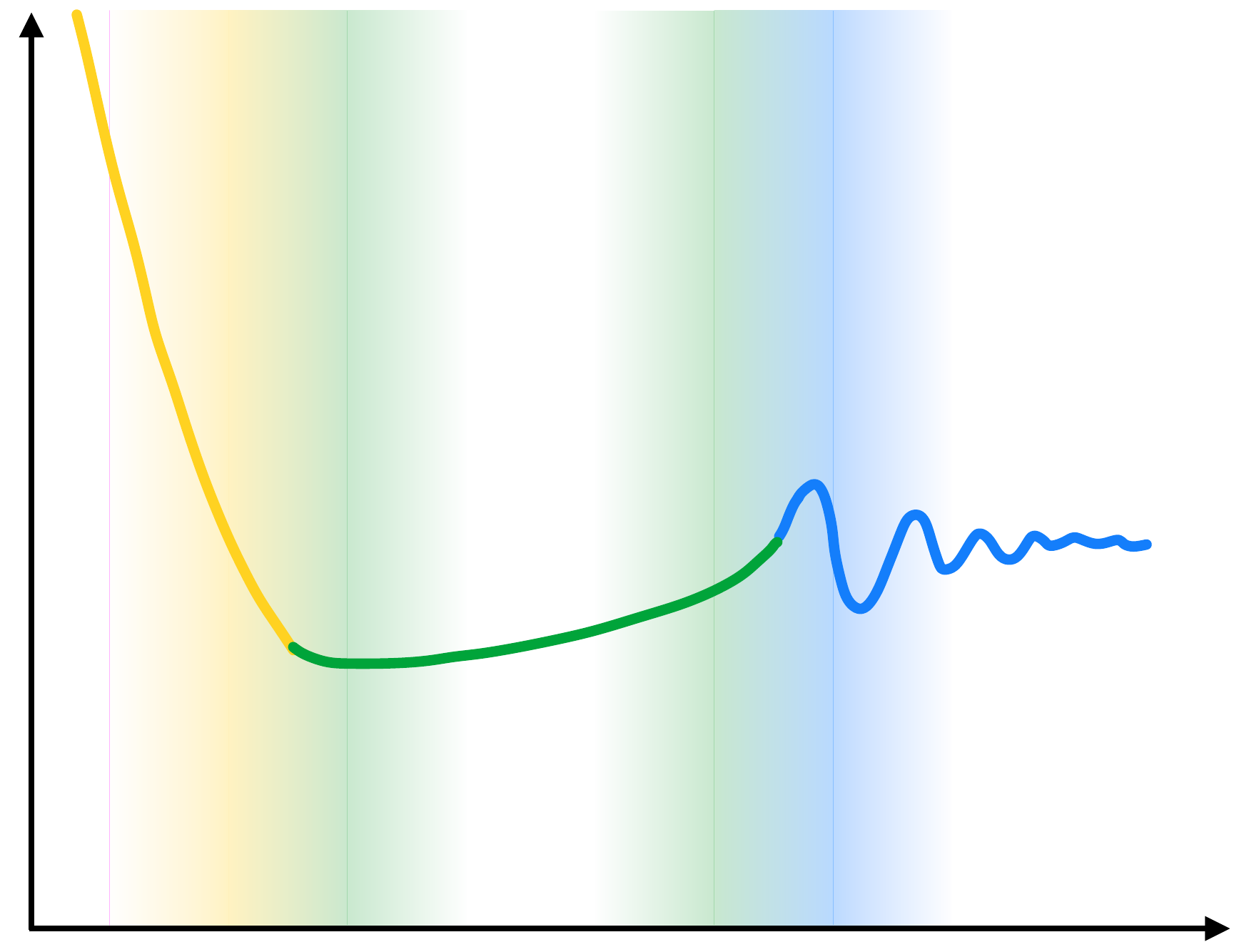}};
\begin{scope}[
x={($0.05*(image.south east)$)},
y={($0.05*(image.north west)$)}]
\node at (10,21) {Physical mode frequency $\omega_k$ for $m_\chi>\tfrac 32 \HI$ and $k<\kend$};
\node at (21,.5) {time};
\node[below] at (12.5,0.5) {$t_\text{end}$};
\draw[gray,thick] (12.5,0.5) -- (12.5,8.6);
\draw[gray,thick] (4.5,0.5) -- (4.5,6.5);
\node[below] at (4.5,0.5) {$t_k$};
\node[above,inner sep=4pt] at (7,0.5) {Inflation};
\node[above,inner sep=2pt] at (15,0.5) {Preheating};
\node[left] at (0.5,5.8) {$\nu$};
\draw[gray] (0.5,5.8) -- (12.5,5.8);
\node[left] at (0.5,8.6) {$m_\chi$};
\draw[gray] (0.5,8.6) -- (18.5,8.6);
\begin{small}
\node[align=center] at (5,18) {Inflationary prod.\\ $\sim$ horizon crossing};
\node[align=center] at (15,18) {Post-infl.~prod.\\ at preheating};
\node[fill=Goldenrod!70!Dandelion, rounded corners=3mm, fill opacity=0.5, text opacity=1] at (3.5,13) {$\omega_k \simeq \frac{k}{a} \sim e^{-H t}$};
\node[fill=ForestGreen!70!Green, rounded corners=3mm, fill opacity=0.3, text opacity=1] at (8.5,4.4) {$\omega_k \simeq \sqrt{m_\chi^2 - \frac 94 H^2}$};
\node[fill=RoyalBlue!90!Blue, rounded corners=3mm, fill opacity=0.2, text opacity=1] at (16.5,5.7) {$\omega_k \simeq \sqrt{m_\chi^2 + \frac 94 w H^2}$};
\end{small}
\end{scope} 
\end{tikzpicture}
\vspace{-1em}
\caption{Evolution of the physical mode frequency $\omega_k$ defined in \cref{eq:nu_def,eq:omega_k} as a function of cosmic time $t$, for a massive scalar ($m_\chi >\HI$) and $k<\kend =\aend \Hend$.
The zeros (in the complex $t$ plane) of the analytical extension of $\omega_k(t)$ determine the relevant epochs of particle production, and the saddle points (see Fig.~\ref{fig:poles}) for the evaluation of the Bogoliubov coefficient $\beta_k$.
Until the time $t_k$ (fixed by $\tfrac{k^2}{a(t_k)^2} = m_\chi^2 - \tfrac 94 H(t_k)^2$, just before Hubble exit), the dispersion relation is dominated by the exponentially decaying physical momentum (in yellow), while during the green epoch $\omega_k$ is almost a constant. 
After the end of inflation $\tend$, the $a(t)$-dependent term oscillates as an effect of the pressure variation during the inflaton oscillations around its minimum (preheating).
Particle production of the massive scalar occurs around the transition regions of Hubble crossing and preheating.
}
\label{fig:omega_zeroes}
\end{figure}

In our paper, we present a comprehensive analytical and numerical analysis of heavy particle production during and after inflation.
To compute particle production rates during these phases,
we employ the Wentzel-Kramers-Brillouin (WKB) approximation and the steepest descent method~\cite{Chung:1998bt, Enomoto:2013mla, Enomoto:2020xlf} to determine the Bogoliubov coefficients.
Unlike the Stokes line method utilized in previous studies about gravitational production 
\cite{Li:2019ves, Li:2020xwr,Enomoto:2020xlf, Hashiba:2022bzi},
the steepest descent method simplifies the treatment
as it eliminates the need to compute Stokes multipliers~\cite{Hashiba:2022bzi}. 
Using the WKB approximation and steepest descent method, we can keep track of particle production around the two transition regions at the times of Hubble crossing and preheating, as illustrated in \cref{fig:omega_zeroes}.
For our numerical analysis, we solve the equations of motion of $\chi$, evaluating the particle occupation number at late times.
We conduct a thorough comparison of the occupation number computed with analytic and numerical approaches.
Finally, we relate the produced number density to the present abundance of the scalar, to assess its viability as a dark matter candidate.

Our study encompasses three distinct inflationary scenarios. 
We first consider a simple de Sitter space transitioning smoothly to a flat Minkowski background, to recover previous findings of the exponentially suppressed particle production.
In the second scenario, we explore the analytic model power-law inflation \cite{Abbott:1984fp, Lucchin:1984yf} to examine particle production with a varying Hubble rate during inflation. 
To understand the dependence on the inflationary model, and compare the particle production occurring during and after inflation, we consider as a third scenario the Starobinsky model of inflation~\cite{Starobinsky:1980te}, which aligns well with current data from Planck and \textsc{BICEP}. 

This paper is structured as follows. 
In \cref{sec:gpdarkmatter}, we discuss the gravitational production of spectator scalar fields, introducing the Bogoliubov transformation and 
the method of steepest descent. 
\Cref{sec:models} examines three specific inflationary scenarios: the transition from a de Sitter phase to a flat Minkowski universe, power-law inflation, and the Starobinsky model of inflation. 
We compute the dark matter abundance for these models in \cref{sec:dmabund}. 
Our conclusions are presented in Section~\ref{sec:conclusions}. 
Additionally, a detailed discussion of the higher-order WKB approximation and a comparison between the steepest descent method and the Stokes line approach are provided in \cref{app:wkbapproximation} and in \cref{app:stokes}. 

\section{Gravitational Production of a Spectator Field}
\label{sec:gpdarkmatter}

In this section, we establish the foundation for evaluating the particle number density resulting from gravitational production.
The production rate is determined by solving the field equations and employing WKB approximation.
We then compute the particle occupation number at the late time of the universe using the Bogoliubov transformation approach,
with the Bogoliubov coefficients derived through the steepest descend approach.

\subsection{Bogoliubov Transformation}
To compute particle production in curved spacetime, we employ the Bogoliubov 
transformation approach~\cite{Parker:1969au, Zeldovich:1971mw, Kofman:1997yn}.  This method allows us to describe the adiabatic evolution 
of the produced particle wave function, taking into account the effects of curved spacetime. The evolution results in
a mixture of positive and negative frequency modes, which characterizes the particles generated by the expansion of spacetime.

We begin our discussion by considering a background geometry that is homogeneous and isotropic, and 
can be effectively modeled using the Friedmann-Robertson-Walker (FRW) spacetime metric:
\begin{equation}
    ds^2 \; = \;  g_{\mu \nu } d x^\mu \, d x^\nu = \, dt^2 - a(t)^2 d \mathbf{x}^2 \, ,
    \label{eq:frwmetric}
\end{equation}
where $t$ represents cosmic time, $a(t)$ is the dimensionless scale factor, and $\bf{x}$ denotes the $3$-dimensional comoving spatial vector. 
Throughout this paper, we focus on cosmic rather than conformal time. 

We introduce a spectator scalar field, $\chi$, with an action given by 
\begin{equation}
    \label{eq:spectatoraction}
    \mathcal{S}_{\chi} \; = \; \int \d^4 x \sqrt{-g} \left(\frac{1}{2}g^{\mu \nu} \partial_{\mu} \chi \partial_{\nu} \chi 
         - \frac{1}{2}m_{\chi}^2 \chi^2 
      \right) \, .
\end{equation}
The kinetic term of the field is canonically normalised by introducing a rescaled
field $X$,\footnote{Note that rescaling the scalar field introduces ambiguity in determining particle production density via the Bogoliubov method, as it may incorporate local contributions to the energy density. This ambiguity can be resolved by measuring the particle number in (or near) flat spacetime.} 
\begin{equation}
   \label{eq:field rescaling}
   X (t, {\bf x}) \; \equiv \; a(t)^{\frac{3}2} \, \chi (t, {\bf x}) \, .
\end{equation}
With this redefinition, the above action is modified in
\begin{equation}
    \label{eq:spectatoractionX}
    \mathcal{S}_X  \; = \; \int \d^4 x  {\cal L}_X \; = \; \int \d^4 x  \left( 
      \frac{1}{2}g^{\mu \nu} \partial_{\mu} X (t, {\bf x}) \partial_{\nu} X  (t, {\bf x})
         -\frac{1}{2} H^2(t)\, \nu^2(t)\,  X^2 (t, {\bf x}) 
      \right) \, ,
\end{equation}
where $H = \dot a / a$ is the Hubble rate, and we define the dimensionless variable,
\begin{equation}
\label{eq:nu_def}
   \nu^2 \; = \; 
   \frac{m_{\chi}^2}{H^2} - \frac{9}{4} - \frac{3}{2} \frac{\dot{H}}{H^2} =
   \frac{m_{\chi}^2}{H^2} + \frac{9}{4} w \, ,
\end{equation}
where $w(t)=p(t)/\rho(t)$ is the equation of state of the Universe.
The field $X$ can be expressed in Fourier space as
\begin{equation}
    \label{eq:fourdecomp}
   X (t, \mathbf x) \; = \; 
      \int \frac{d^{3} \mathbf{k}}{ {(2 \pi )^{3}}} \, 
      \left[\hat{a}_{\bf k}^{\,} X_k (t)  \, e^{i {\bf k} \cdot {\bf x} }
       +  \hat{a}^\dagger_{\bf k} X_k^* (t)  \, e^{ - i {\bf k} \cdot {\bf x} } 
      \right] \, .
\end{equation}
Here $\bf{k}$ denotes the comoving momentum vector, and $\hat{a}^{\dagger}_{\bf k}$ and $\hat{a}_{\bf k}^{\,}$ are the creation and annihilation operators, 
respectively, that satisfy the commutation relations $[\hat{a}_{\bf k}, \hat{a}_{\bf k'}] = [\hat{a}_{\bf k}^{\dagger}, \hat{a}_{\bf k'}^{\dagger}] = 0$
and $[\hat{a}_{\bf k}^{\,}, \hat{a}_{\bf k'}^{\dagger}] = (2 \pi )^3\delta^{3}(\mathbf{k} - \mathbf{k}')$. 
From the action given in Eq.~(\ref{eq:spectatoractionX}), we derive the conjugate momentum of $X$, given by 
$\pi \equiv \partial \mathcal{L}_{X}/\partial \dot{X} = \dot{X}$. 
To quantize the system, we impose the commutation relation
\begin{equation}
    [X(t, \mathbf{x}), \pi(t, \mathbf{y})] \; = \; i \delta^{3}(\mathbf{x} -\mathbf{y}) \, ,
\end{equation}
which, along with the commutation relations of the creation and annihilation operators, implies the Wronskian condition:
\begin{equation}
    \label{eq:wronskian}
    X_k \dot{X}_k^* - X_k^*\dot{X}_k \; = \; i \, .
\end{equation}
The field equation for $X$ in Fourier space, as a consequence of the rescaling in Eq.~\eqref{eq:field rescaling}, does not display a term of Hubble friction, and is given by
\begin{equation}
    \label{eq:modeeq1}
    \ddot{X}_k(t) + \omega_k^2(t) X_k(t) \; = \; 0 \, ,
\end{equation}
with
\begin{equation}
\label{eq:omega_k}
\omega_k^2 \; = \; \frac{k^2}{a^2} + H^2 \nu^2 \,  .
\end{equation}
Next, applying the \textit{adiabatic} (WKB) approximation, we assume that the wave function $X_k(t)$ can be expressed as the sum of its solutions $\exp(\pm i \int \omega_k dt)$~\cite{Kofman:1997yn}
\begin{equation}
    \label{eq:wavefunct}
    X_k(t) \; = \; \frac{\alpha_k}{\sqrt{2\omega_k}} e^{-i \int^{t} \omega_k d t} + \frac{\beta_k}{\sqrt{2 \omega_k} }e^{+i \int^{t} \omega_k d t} \, ,
\end{equation}
where $\alpha_k$ and $\beta_k$ are the time-dependent Bogoliubov coefficients. 

The adiabatic approximation is exact in the limit in which the wave function evolves adiabatically and satisfies the condition 
$|\dot{\omega}_k| \ll |\omega_k|^2$ at all times during the evolution.
In Eq.~\eqref{eq:wavefunct}, $\omega_k$ corresponds to the zeroth-order of the WKB expansion,
while high-order frequency will be implemented in the numerical analysis to achieve a faster convergence of the results to the actual solution. 
The frequency for the $j$-th order improved WKB approximation is given by the iterative formula,
\begin{equation}
\omega_k^{(j)} \; = \; 
\sqrt{\omega_k^2-\left[\frac{\ddot{\omega}_k^{(j-1)}}{2 \omega_k^{(j-1)}}-\frac{3}{4}\left(\frac{\dot{\omega }_k^{(j-1)}}
      {\omega_k^{(j-1)}}\right)^2\right]} \, ,
   \label{eq:omega_jth}
\end{equation}
starting with $\omega_k^{(0 )}  = \omega_k$. Further discussion on the high-order WKB approximation is provided in Appendix~\ref{app:wkbapproximation}.

The time-dependent Bogoliubov coefficients are correlated via the field equations \cref{eq:modeeq1}, leading to the following relationships:
\begin{equation}
    {\dot \alpha}_k(t) \; = \; \frac{\dot\omega_k}{2\omega_k} \beta_k  e^{ 2 i \int^{t} \omega_k d t}
\,, \qquad~{\dot \beta}_k(t) \; = \;  \frac{\dot \omega_k}{2\omega_k} \alpha_k    e^{ - 2 i \int^{t} \omega_k d t}
\, .
\end{equation}
By combining these equations with the Wronskian condition in Eq.~(\ref{eq:wronskian}), we find the normalization condition for the Bogoliubov coefficients:
\begin{equation}
    \label{eq:bogoliubov}
|\alpha_k|^2 - |\beta_k|^2 \; = \; 1 \, .
\end{equation}
To solve the mode equation~(\ref{eq:modeeq1}), we must impose the initial conditions. Using the WKB approximation in the early-time limit, $t \rightarrow -\infty$, we choose the well-known 
Bunch-Davies conditions for the initial vacuum state:
\begin{equation}
    \lim_{t_i \rightarrow - \infty} X_{k} (t_i) \; = \; \frac{1}{\sqrt{2 \omega_k ( t_i)}} e^{-i  \int^{t_i} \omega_k  d t} \, .
   \label{eq:BDvac}
\end{equation}
Comparing this to the adiabatic form of the wave function in Eq.~(\ref{eq:wavefunct}), one sees that
\begin{equation}
    \label{eq:initialbogoliubov}
    \alpha_k(t_i) \; = \; \, 1\, , \qquad \beta_k(t_i) \; = \; 0 \, ,
\end{equation}
where $t_i$ is an initial time satisfying the Bunch-Davies vacuum conditions.

The average energy density of the spectator field at any given time, $ \langle H(t) \rangle $, is evaluated as
\begin{equation} 
\langle H(t) \rangle =  
\left\langle \Omega \left|  
  \int \d^3 {\bf x}  \, \Big( \pi(t, {\bf x})  X ( t,{\bf x} ) - {\cal L}_X \Big) \, 
\right| \Omega \right\rangle =
  \int \frac{\d^3 {\bf k} }{ ( 2 \pi )^3} \frac{1}{2} \left(|{\dot X}_k |^2 +  \omega_k^2  | {X}_k |^2 
\right) .
\end{equation} 
We define the dimensionless occupation number $n_k$ as the energy density for a mode $k$ divided by the frequency:
\begin{equation}
    n_k(t) \;  = \; \frac{1}{\omega_k} \left(\frac 12|\dot{X}_k(t)|^2 +\frac 12\omega_k^2 |X_k(t)|^2 \right) - \frac{1}{2} \; 
         =  \frac{1} {2 \omega_k}   \left|  {\dot X}_k+ i \omega_k  X_k \right|^2   \simeq \; |\beta_k|^2 \, ,
   \label{eq:n_k}
\end{equation}
where the leading vacuum energy contribution $ -\frac{1}{2}$ is removed. 
The second equality in Eq.~\eqref{eq:n_k} is derived using the Wronskian condition, and the approximate equality is validated by employing the WKB approximation in Eq.~(\ref{eq:wavefunct}).
It is important to note that the normal-ordering subtraction of $\tfrac 12$ in Eq.~\eqref{eq:n_k} does not eliminate all local contributions to the energy density.
The previous definition may leave an ambiguity in the particle occupation number at late times.
However, if the inflationary background is connected to a Minkowski or matter-dominated universe, $n_k(t)$ corresponds to the occupation number of produced particles as $t \to \infty$, since the geometric contribution will be subdominant compared to the particle energy density. 
In numerical approaches, this quantity is computed at sufficiently late times. 

In the numerical results shown in the paper, we solve for $X_k(t)$ using Eq.~\eqref{eq:modeeq1}, and compute 
$n_k$ by replacing $\omega_k$ in Eq.~\eqref{eq:n_k} with the higher-order WKB approximation $\omega_k^{(j)}$.
The occupation number $n_k$ carries a small dependence on the order $(j)$, stemming from the ambiguity of the vacuum states in curved spacetime or from additional vacuum contributions that necessitate subtraction. 
These differences vanish when evaluating the expression for $n_k$ in Minkowski spacetime, or at asymptotically late times $\tfinal \gg \tend$, so that the particle occupation number becomes independent of the order of the WKB approximation. See Refs.~\cite{Dabrowski:2014ica, Dabrowski:2016tsx} for a detailed discussion.

The physical number density is obtained by integrating the occupation number $n_k$,
\begin{equation}
    \label{eq:number_density}
    n_{\chi} \; = \; \frac{1}{2 \pi^2 a(t)^3} \int_{0}^{\infty} dk \, k^2 |\beta_k|^2 \, .
\end{equation}
Gravitational particle production in the regime $|\beta_k|^2 \ll 1$, $ |\alpha_k| \simeq 1$, leads to~\cite{Kofman:1997yn} \begin{equation}
\label{eq:bogbetait}
\beta_k(t) \simeq \frac{1}{2} \int_{t_i}^{t}\d t' \,  \frac{\dot{\omega}_k}{\omega_k} \exp\Big(-2i \Omega_k(t')\Big) \, ,\quad
\Omega_k(t) \equiv \int_{t_i}^{t} \d t' \, \omega_k(t') \, .
\end{equation}
We now explore analytical methods to estimate this solution for $\beta_k$.

\subsection{Steepest Descent Method and Particle Number Density}
\label{sec:steepest}

After having introduced the Bogoliubov transformation to derive the occupation number in Eq.~\eqref{eq:bogbetait}, we proceed to evaluate the integral using the steepest descent method, paralleling the methodology outlined in \cite{Chung:1998bt, Enomoto:2013mla, Enomoto:2020xlf}.

The steepest descent method shows that the dominant contribution to the integral in Eq.~\eqref{eq:bogbetait} arises from the region near the saddle points in the complex $t$ plane.
We deform the integration contour for $\beta_k$ so that it passes through these saddle points along the appropriate directions, where the real part of the exponential term in the integrand decreases most rapidly and the imaginary part is constant (hence the alternative name of ``constant phase''). 
This guarantees that the integrand decreases quickly as one moves away from the saddle point.

The first step is to identify the saddle points for the integral in \cref{eq:bogbetait}. They are determined by setting the derivative of the exponent to zero, $\frac{\d}{\d t} \Omega (t) = 0$, which leads to 
\begin{equation} 
   \omega_k (t_n) \; = \; 0 \,.
\end{equation} 
These saddle points for the integral of $\beta_k$ happen to coincide with the \textit{poles} of the integrand, and we often refer to them as $n$-th pole.
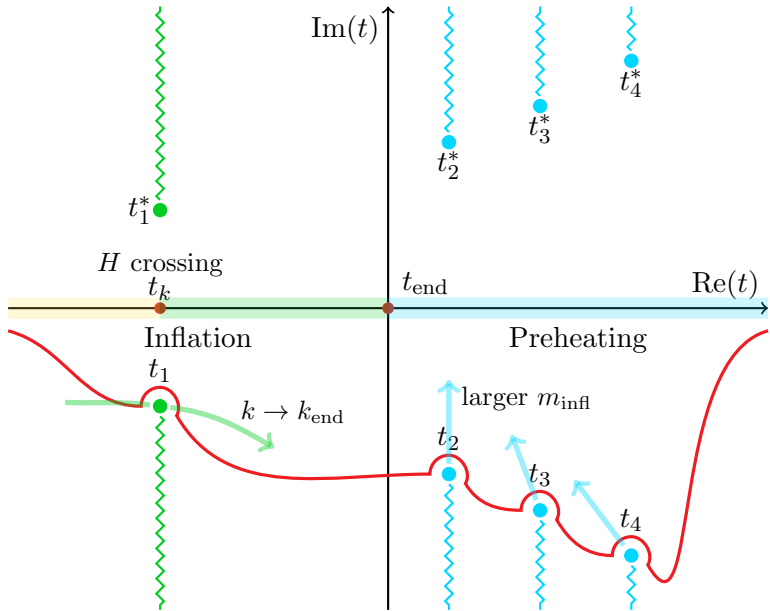
\begin{figure}\centering
Integration contour for $\beta_k$ in the steepest descent approximation \smallskip\\
\begin{tikzpicture}
\def\rp{.25} 
\def\tlab{9pt} 
\def\x{5} \def\y{4} 
\def\dtRH{1.2} 
\draw[thick,->] (-\x,0) -- (\x,0);
\draw[thick,->] (0,-\y) -- (0,\y);
\draw [Red,fill=BrickRed] (0,0) circle [radius=2.pt];
\node[above left,inner sep=3pt] at (\x,0){$\Re(t)$};
\node[below left,inner sep=3pt] at (0,\y){$\Im(t)$};
\node[above right,inner sep=5pt] at (0,0){$t_\mathrm{end}$};
\node[below,inner sep=7pt] at (-0.5*\x,0){Inflation};
\node[below,inner sep=7pt] at (0.5*\x,0){Preheating};
\node at (-3,-1.3)(t1){};
\node[above,inner sep=\tlab] at (t1){$t_1$};
\draw [poleCross,fill=poleCross] (t1) circle [radius=2.5pt];
\draw[decoration = {zigzag,segment length = 2mm, amplitude = .5mm},decorate, poleCross,thick] (t1)--($(-\x,-\y)!(t1)!(0,-\y)$);
\node[above,inner sep=3pt] at ($(-\x,0)!(t1)!(0,0)$){$t_k$};
\draw [Red,fill=BrickRed] ($(-\x,0)!(t1)!(0,0)$) circle [radius=2.pt];
\draw[line width=8pt,opacity=.2,Goldenrod] (-\x,0) -- ($(-\x,0)!(t1)!(0,0)$);
\draw[line width=8pt,opacity=.2,poleCross] ($(-\x,0)!(t1)!(0,0)$) -- (0,0);
\draw[line width=8pt,opacity=.2,poleRH] (0,0) -- (\x,0);
\node at (.8,-2.2)(t2){};
\node[above,inner sep=\tlab] at (t2){$t_2$};
\draw[poleRH,fill=poleRH] (t2) circle [radius=2.5pt];
\draw[decoration = {zigzag,segment length = 2mm, amplitude = .5mm},decorate,poleRH, thick] (t2)--($(-\x,-\y)!(t2)!(0,-\y)$);
\node at ($(t2)+(\dtRH,-0.4*\dtRH)$)(t3){}; 
\node[above,inner sep=\tlab] at (t3){$t_3$};
\draw[poleRH,fill=poleRH] (t3) circle [radius=2.5pt];
\draw[decoration = {zigzag,segment length = 2mm, amplitude = .5mm},decorate,poleRH, thick] (t3)--($(-\x,-\y)!(t3)!(0,-\y)$);
\node at ($(t3)+(\dtRH,-0.5*\dtRH)$)(t4){}; 
\node[above,inner sep=\tlab] at (t4){$t_4$};
\draw[poleRH,fill=poleRH] (t4) circle [radius=2.5pt];
\draw[decoration = {zigzag,segment length = 2mm, amplitude = .5mm},decorate,poleRH, thick] (t4)--($(-\x,-\y)!(t4)!(0,-\y)$);
\draw[very thick,Red] (-\x,-0.3) 
    to [out=-15, in=180] ($(t1)+(-\rp,0*\rp)$)
    to [out=90, in=180] ($(t1)+(0,\rp)$)
    to [out=0, in=60]($(t1)+(.87*\rp,-.5*\rp)$)
    to [out=-60, in=180] ($(t2)+(-\rp,-0*\rp)$)
    to [out=90, in=180] ($(t2)+(0,\rp)$)
    to [out=0, in=60]($(t2)+(.87*\rp,-.5*\rp)$)
    to [out=-60, in=180] ($(t3)+(-\rp,-0*\rp)$)
    to [out=90, in=180] ($(t3)+(0,\rp)$)
    to [out=0, in=60]($(t3)+(.87*\rp,-.5*\rp)$)
    to [out=-60, in=180] ($(t4)+(-\rp,0*\rp)$)
    to [out=90, in=180] ($(t4)+(0,\rp)$)
    to [out=0, in=60]($(t4)+(.87*\rp,-.5*\rp)$)
    to [out=-60, in=-165](\x,-0.3);
\draw[line width=2pt,opacity=.4,poleCross,->] ($(t1)+(-5*\rp,.2*\rp)$) 
    to [out=-2, in=173] (t1)
    to [out=-7, in=150] ($(t1)+(6*\rp,-2.2*\rp)$);
\node[above,inner sep=\tlab] at ($(t1)+(7*\rp,-2.2*\rp)$){\small $k\to k_\mathrm{end}$};
\draw[line width=2pt,opacity=.4,poleRH,->] (t2) 
    to [out=90, in=-90] ($(t2)+(0,5*\rp)$);
\node[right,inner sep=.5*\tlab] at ($(t2)+(0,4*\rp)$){\small larger $m_\mathrm{infl}$};
\draw[line width=2pt,opacity=.4,poleRH,->] (t3) 
    to ($(t3)+(-1.5*\rp,4*\rp)$);
\draw[line width=2pt,opacity=.4,poleRH,->] (t4) 
    to ($(t4)+(-3*\rp,4*\rp)$);
\begin{scope}[yscale=-1,xscale=1]
\def\tlab{3pt} 
\node at (-3,-1.3)(t1){};
\node[left,inner sep=\tlab] at (t1){$t_1^*$};
\node[above,inner sep=4*\tlab] at ($(-\x,0)!(t1)!(0,0)$){\small $H$ crossing};
\draw [poleCross,fill=poleCross] (t1) circle [radius=2.5pt];
\draw[decoration = {zigzag,segment length = 2mm, amplitude = .5mm},decorate, poleCross,thick] (t1)--($(-\x,-\y)!(t1)!(0,-\y)$);
\node at (.8,-2.2)(t2){};
\node[below,inner sep=\tlab] at (t2){$t_2^*$};
\draw[poleRH,fill=poleRH] (t2) circle [radius=2.5pt];
\draw[decoration = {zigzag,segment length = 2mm, amplitude = .5mm},decorate,poleRH, thick] (t2)--($(-\x,-\y)!(t2)!(0,-\y)$);
\node at ($(t2)+(\dtRH,-0.4*\dtRH)$)(t3){}; 
\node[below,inner sep=\tlab] at (t3){$t_3^*$};
\draw[poleRH,fill=poleRH] (t3) circle [radius=2.5pt];
\draw[decoration = {zigzag,segment length = 2mm, amplitude = .5mm},decorate,poleRH, thick] (t3)--($(-\x,-\y)!(t3)!(0,-\y)$);
\node at ($(t3)+(\dtRH,-0.5*\dtRH)$)(t4){}; 
\node[below,inner sep=\tlab] at (t4){$t_4^*$};
\draw[poleRH,fill=poleRH] (t4) circle [radius=2.5pt];
\draw[decoration = {zigzag,segment length = 2mm, amplitude = .5mm},decorate,poleRH, thick] (t4)--($(-\x,-\y)!(t4)!(0,-\y)$);
\end{scope}
\end{tikzpicture}
\vspace{-1em}
\caption{The integration path for $\beta_k$ on the complex $t$ plane. The colour shadings on the real axis reflect the colour code of Fig.~\ref{fig:omega_zeroes} about the sub-Hubble, super-Hubble and post-inflationary epochs for the mode frequency $\omega_k(t)$.
The saddle points $t_n$ (given by $\omega_k(t_n)=0$) of the integrand of $\beta_k$ are indicated by dots, and the branch cuts are shown with zigzag lines.
The red line is the contour for the saddle-point approximation of $\beta_k$, whose path around the poles covers an angle $4\pi/3$ avoiding the branch cut.
From a quantitative point of view, the most relevant saddle point for the computation of $\beta_k$ is the closest one to the real $t$ axis.
The green and blue arrows show the displacement of the saddle points as we vary $k$ and the inflaton mass. 
If we increase the spectator field mass $m_\chi$, all saddle points get further away from the real axis, suppressing exponentially the particle production.
}
\label{fig:poles}
\end{figure}

Starting from the squared mode frequency in Eq.~\eqref{eq:omega_k}, we expand $\omega_k^2(t)$ near the $n$-th saddle point to find the steepest descent paths, 
\begin{equation}
\label{eq:Omega approx}
\Omega_k (t) \simeq 
  \int_{t_i}^{t_n} \hspace{-3pt}\d t'\omega_k(t') 
  + \sqrt{g(t_n)} \int_{t_n}^{t} \d t' \sqrt{(t'-t_n)} =
  \int_{t_i}^{t_n} \hspace{-3pt}\d t' \omega_k(t') 
  + \frac{2}{3} \sqrt{g(t_n)}(t - t_n)^{3/2} \, ,
\end{equation}
where
\begin{equation}
g(t_n) \; \equiv \; \frac{\d }{\d t} \omega_k^2(t) \bigg|_{t=t_n}\, . 
\end{equation}
Notice that the square root in \cref{eq:Omega approx} introduces branch cuts originating from the saddle points $t_n$.
With this expansion we approximate the prefactor in \cref{eq:bogbetait} as
\begin{equation}
\frac{\dot{\omega}_k(t)}{2\omega_k(t)} \;\simeq \; \frac{1}{4} \frac{1}{t-t_n} \, .
\end{equation}
We illustrate the steepest descent path of integration on the complex plane in \cref{fig:poles}. 
We deform the contour in the lower half-plane, and \cref{eq:bogbetait} transforms into a sum of the saddle point contributions
\begin{gather}
\label{eq:betaapprox1}
\beta_k \; \simeq \; \sum_n V_n \exp\left[-2i \int_{t_i}^{t_n} \d t' \omega_k(t') \right] \, \\
V_n \; = \; \frac{1}{4} \int_{C_n} \frac{\d t}{t-t_n} \exp \left[-\frac{4i}{3}\sqrt{g(t_n)}(t-t_n)^{3/2} \right] \, .
\end{gather}
We evaluate the coefficients $V_n$ along the deformed contour $C_n$ that approaches the $n$-th pole along the path of steepest descent and then goes around this pole. 
Importantly, this general expression accounts for contributions from all $n$ poles (saddle points). 

The constant-phase paths are selected by the phase $\arg(V_n)=(2j+1)\pi,\, j\in \mathbb Z$. 
Denoting $\theta \equiv \arg(t-t_n),\, \varphi \equiv \arg(g(t_n))$,
\begin{equation}
\label{eq:steepestphase}
\frac{3\pi}{2} + \frac{\varphi}{2}  + \frac{3 \theta}{2} = (2j+1) \pi  \, \Rightarrow \, \theta = \frac{(4j-1)\pi - \varphi}{3} \, ,
\end{equation}
For the dispersion relation of Eq.~\eqref{eq:omega_k} and the supermassive case $m_\chi>\tfrac 32 \HI$ that we consider, $\varphi=0$ for all saddle points. 
Regarding the branch cuts introduced in each saddle point by the square root of Eq.~\eqref{eq:Omega approx}, we choose to align them with vertical lines to $\pm i \infty$ as shown in \cref{fig:poles}. 
The contour $C_n$ can then avoid the branch cut by picking $j=1$ as ingoing phase ($\theta = \pi$) and 
$j=0$ as outgoing phase ($\theta = -\tfrac{\pi}{3}$), as shown in Fig.~\ref{fig:poles}.
Consequently, by evaluating the prefactor along the contour $C_n$ in the limit of vanishing radius, we obtain
\begin{equation}
V_n \; \simeq \; \frac{1}{4} \times \frac{4\pi}{3} i \; = \; \frac{i \pi}{3} \, .
\end{equation}
Next, we evaluate the exponent $\Omega_k(t)$ in \cref{eq:betaapprox1}. 
The contour of integration for $\Omega_k$ can be freely modified, as $\omega_k(t)$ is an analytic function.
It is important to note that the integration path of \cref{eq:betaapprox1} does not need to coincide with the contour for $\beta_k$. We split the integral as
\begin{equation}
    \int_{t_i}^{t_n} dt' \omega_k(t') \; = \; \int_{t_i}^{\Re t_n} dt' \omega_k(t') + \int_{\Re t_n}^{t_n} dt' \omega_k(t') \, .
    \label{eq:Omega_split}
\end{equation}
The first term in \cref{eq:Omega_split} is real, contributing as a phase to each pole's contribution. This term cancels when considering the contribution from a single pole, though it may have a small effect when summing multiple poles, even if the integral along the real axis typically contributes less. 
Therefore, the occupation number for the mode $k$ takes the form
\begin{equation}
\label{eq:saddlepoint1}
|\beta_k|^2 \; \simeq \;  \frac{\pi^2}{9}  \left| \sum_n e^{2 \Im \Omega_k (t_n)} \right|^2
\simeq  \frac{\pi^2}{9} 
\left| \sum_n \exp\left( -2 i  \int_{\Re t_n}^{t_n} \d t' \omega_k(t')  \right)\right|^2\, .
\end{equation}
We use this equation to compute the particle number both analytically and numerically in \cref{sec:models}.
The particular contour for $\Omega_k(t)$ in \cref{eq:Omega_split} is utilized for the models of de Sitter and Starobinsky inflation, while different paths are chosen for power-law inflation to obtain an analytic formula.

The result can further approximated by noting that $\omega_k(t)$ has an almost constant modulus for most of the path from $\mathrm{Re}(t_n)$ to $t_n$. 
Within that approximation,
\begin{equation}
\label{eq:approxwkbint1}
\int_{t_i}^{t_n} \d t' \omega_k(t') \simeq i\, \Im \left( \int_{\Re t_n}^{t_n} \d t' \omega_k(t') \right)
 \simeq i\,   \omega_k \Big(\Re(t_n)\Big) \times \Im(t_n) \,,
\end{equation}
and the occupation number of produced particles becomes 
\begin{equation}
\label{eq:saddlepointapprox}
|\beta_k|^2 \; \simeq \;  \frac{\pi^2}{9} \left| \sum_n e^{2 \omega_k\left(\Re(t_n)\right) \times \Im(t_n)} \right|^2\, .
\end{equation}
This result provides a good analytical estimate. 

Alternatively, Stokes's method can be employed for the integration. The comparison between 
the steepest descent method and Stokes's method is detailed in \cref{app:stokes}.

\section{Inflationary Models}
\label{sec:models}

In this section, we explore the model dependence of gravitational particle production, examining its occurrence during and after inflation within a unified framework. 
Our examination encompasses three distinct scenarios: 
de Sitter spacetime, power-law inflation, and the Starobinsky model of inflation. 
To ensure a comprehensive analysis and comparison, we apply both the numerical solution and the analytical steepest descent approach to each scenario.

\subsection{de Sitter to Minkowski universe transition}
\label{sec:dstomink}

We explore a simple scenario where de Sitter spacetime smoothly transitions into Minkowski spacetime, characterized by the scale factor, 
\begin{equation}
a(t)  = 2\aend \frac{e^{\HI(t-t_{\rm{end}})}}{1+ e^{\HI (t - t_{\rm{end}})}} \, .
\end{equation}
Here, $\HI$ represents a constant Hubble parameter during the de Sitter phase and $\tend$ marks the transition time from de Sitter to flat spacetime. 
This schematic cosmological evolution simplifies the calculation of the gravitationally produced abundance because it displays a quick transition to an asymptotically flat spacetime.

\begin{figure}\centering
de Sitter inflation to Minkowski\\
\includegraphics[width=0.5\textwidth]{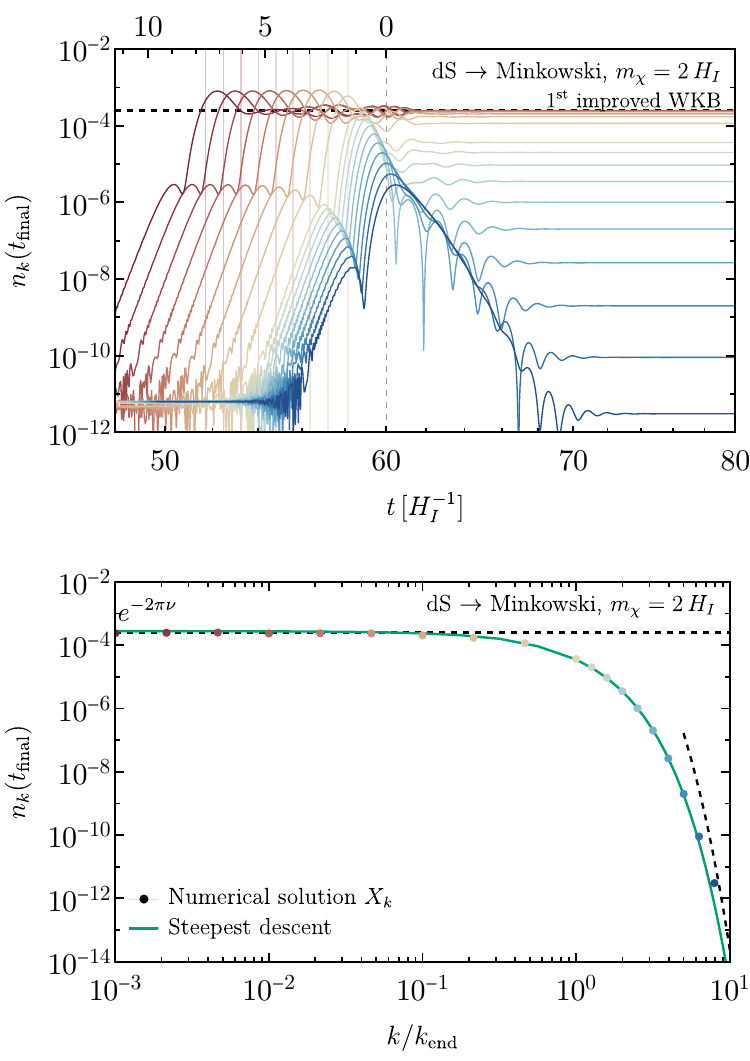}  \hspace{-1em}\hfill
\includegraphics[width=0.5\textwidth]{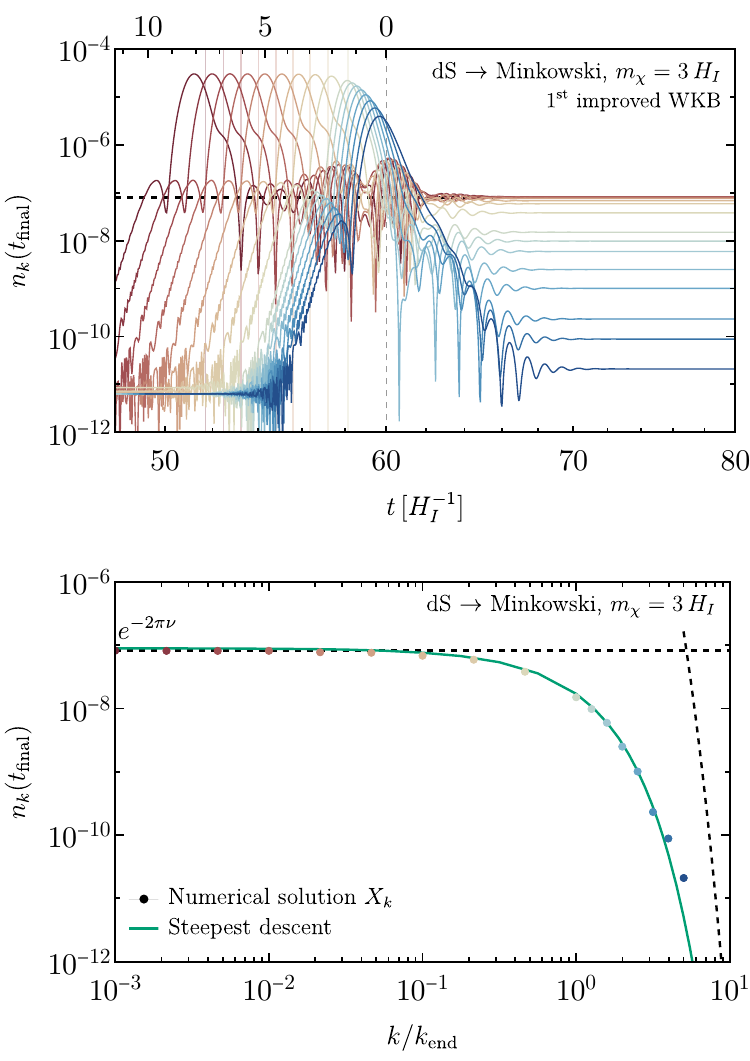}
\caption{Particle occupation number $n_k = |\beta_k|^2$ as a function of time $t$ in $\HI^{-1}$ units (top panels) and of momentum $k/\kend$ (bottom panels, where $n_k$ is evaluated at a late time $\tfinal$) for two representative mass choices of $m_{\chi} = 2\HI$ (left panels) and $m_{\chi} = 3\HI$ (right panels). 
Each colored line in the top plots corresponds to the momentum marked in the bottom plots by a point with the same color. The color grading from red to blue reflects the transition from IR ($k<\kend$) to UV modes ($k>\kend$).
In the bottom plots, the green line is the result of the steepest descent method where the saddle points and the integral of Eq.~\eqref{eq:saddlepoint1} are evaluated numerically.
The black dashed lines show the analytical approximations of the steepest descent method computed in this section.
The steepest descent approximation is in excellent agreement with the fully numerical computation. 
}
\label{fig:1_dS-Minkowski_nk_t-k}
\end{figure}

The following discussion includes the gravitational particle production of a heavy scalar field, analyzed using two complementary approaches: the numerical solution of the scalar field evolution in the spacetime background and the approximate WKB approach. \cref{fig:1_dS-Minkowski_nk_t-k} shows the consistency of the two methods.

\subsubsection*{Numerical approach}
We first determine the plane-wave solutions of the scalar field numerically by solving the field equation
$\ddot{X}_k + \omega_k^2 (t) X_k = 0 $, with 
\begin{equation}
\omega_k^2 = \frac{k^2}{a^2} + \HI^2 \nu^2 =
  \frac{ k^2 ( 1+ e^{\HI (t-\tend )})^2 } {4\aend^2 e^{2\HI (t-\tend )}  }
  + m^2 
  +\frac{3}{4} \frac{ (-3 +  2 e^{\HI (t-\tend)})} {( 1+ e^{\HI (t-\tend )}  )^2 } \HI^2  \, . 
\label{eq:omegak_dS}
\end{equation}
The expression for $\omega_k^2$ in the specific spacetime background is detailed in the equation. The initial condition is set at a sufficiently early time using the WKB approximation, as expressed in \cref{eq:BDvac}.
We then solve the field equation for $X_k(t)$ with the initial condition.
Following this, the occupation number $n_k$ at a given time $t$ (\ref{eq:n_k})
is evaluated using the mode frequency for the $j$-th order improved WKB approximation shown in \cref{eq:omega_jth}.
In \cref{fig:1_dS-Minkowski_nk_t-k}, we use the first-order improved WKB approximation for a better numerical precision.
The particle occupation number $n_k = |\beta_k|^2$ is presented as a function of $t$ in units of $\HI^{-1}$ (top panels) for two distinct mass choices $m_{\chi} = 2\HI$ (left panels) and $m_{\chi} = 3\HI$ (right panels). 
The time evolution of $n_k$ reveals an initial increase followed by a subsequent decrease. 
This behavior is not physical and changes if we use a different order of WKB approximation: the physical quantity is the asymptotic value reached at late times.
From a numerical perspective, the absolute accuracy in tracking the exponentially suppressed occupation number $n_k$ is set by the initial value of the mode function $|X_k(t_i)| =1/\sqrt{2\omega_k(t_i)} \simeq \sqrt{a(t_i)/k}$. Pushing $t_i$ to earlier times (especially at small $k$) increases the numerical precision of $n_k$, which can be read off in our results from the (unphysical) plateau at early times in the top panels of Figs.~\ref{fig:1_dS-Minkowski_nk_t-k}, \ref{fig:2_Power-Law_nk_t-k}, \ref{fig:3_Starobinsky_a5_nk_t-k}, \ref{fig:3_Starobinsky_a1_nk_t-k}.

The final spectra for the occupation number $n_k$ are evaluated at a late time $\tfinal =200\, \HI^{-1} \gg \tend$ (we compute a time average of the small residual oscillations), and are shown in the bottom panel of \cref{fig:1_dS-Minkowski_nk_t-k} as a function of the rescaled momentum $k/\kend$, where $\kend\equiv \aend \HI$ is the mode crossing the Hubble radius at the end of inflation. 

The spectrum $n_k$ comprises a UV ($k \gg \kend$) and an IR part ($k \ll \kend$). 
The long wavelength (IR) spectrum is flat in $k$, as shown in \cref{fig:1_dS-Minkowski_nk_t-k}, consistent with known results for particle production in de Sitter for heavy scalar fields.
The IR spectrum is $n_k \sim \exp(-2 \pi\nu)$ with $\nu \equiv \sqrt{ m^2 / \HI^2 - 9/4}$, marked by horizontal dashed lines in the bottom panels of \cref{fig:1_dS-Minkowski_nk_t-k}.
This occupation number, in the limit $m_\chi\gg \HI$, is analogous to the Bose-Einstein distribution for a heavy particle at a Gibbons-Hawking temperature $T_\text{dS} = \HI/2 \pi \ll m_\chi $.

The short wavelength (UV) spectrum is exponentially suppressed with respect to $k$. 
These modes never left the Hubble radius during inflation. 
From the numerical perspective of the equations of motion for $X_k$, the non-adiabaticity parameter $\dot \omega_k/\omega_k^2$ is suppressed by the large $\omega_k\sim k/a$ during inflation, reducing the gravitational production for these modes.

\subsubsection*{Analytical approach}
We now compare the numerical results obtained by solving the equations of motion for $X_k$, with the analytical formula for $\beta_k$ obtained with the steepest descent approximation, as detailed in \cref{sec:steepest}.
We first consider the IR modes, for which particle production occurs before the end of inflation. 
In this first model where the inflationary epoch is almost exactly de Sitter, the Hubble rate is a constant $\HI$. 
The saddle points $t_n$ are easily obtained in this case by solving $\omega_k(t_p) = 0$ with constant $\HI$ and $\nu$:
\begin{equation}
\label{eq:dS pole Hcrossing}
\HI ( t_n - \tend ) = \log ( \frac{ k} {\aend\HI \nu } ) -i \frac{ \pi}{2}  - i n \pi \, , \quad n \in \mathbb Z \, .
\end{equation}
The dominant saddle point in the lower half-plane is at $n = 0$, since it is closest to the real $t$ axis and minimizes the exponential suppression in \cref{eq:bogbetait}. 
That formula, for the case of one saddle point, simplifies into
\begin{equation}
n_k \simeq  | \beta_k|^2 \simeq \frac{\pi^2}{9} \exp 
\left( -2 i \int_{t_p^*}^{t_p} \d t \, \omega_k(t) \right) \, .
\label{eq:nk_dS}
\end{equation}
The exponent of this expression, for the pole of \cref{eq:dS pole Hcrossing}, is equal to
\begin{equation}
-2 i \int_{t_p^*}^{t_p} \d t \, \omega_k ( t)  = - 2i \nu \int_{\frac{\pi}{2} i}^{ - \frac{\pi}{2} i} \d (\HI t) 
\sqrt{e^{-2\HI t} +1}  = - 2\pi \nu  \, .
\end{equation}
The occupation number for IR modes is then
\begin{equation}
      n_k^\textsc{(ir)} \; \simeq \; \frac{\pi^2}{9} e^{-2 \pi \nu} \, . 
\end{equation}
This result agrees with alternative methods, such as the in-out formalism~\cite{Anderson:2013ila, Anderson:2013zia, Markkanen:2016aes} and the Stokes line method~\cite{Li:2019ves, Corba:2022ugu}.

Short wavelength modes $k \gg \kend$ never cross the Hubble radius during inflation. 
Particle production for UV modes hence occurs at $\tend$ when the spacetime transitions into Minkowski.
By considering Fig.~\ref{fig:omega_zeroes}, the green band of $\omega_k\sim \HI\nu$ disappears, and the yellow band where $\omega_k\sim k e^{\HI(\tend-t)}/\aend$ directly connects to the post-inflationary phase (shaded in blue).
When solving $\omega_k(t_n)=0$ for the saddle points in the limit $k\gg \kend$ limit, from \cref{eq:omegak_dS} at 0th order in $\tfrac{k}{\kend}$ we see that $1+ e^{\HI (t-\tend )}$ must be very close to 0, so that the gradient term in \cref{eq:omegak_dS} is sufficiently small that the overall expression for $\omega_k^2$ can vanish.
As a result, the poles are around $ t_p \sim \tend + i\pi\HI^{-1}(2n+1),\, n\in \mathbb Z$. 
Solving $\omega_k  =0 $ around $\tend - i\pi \HI^{-1}$ in the limit $k\gg \kend$, yields four roots,
\begin{equation}
e^{\HI (t-\tend )}  = -1 + e^{i\frac{\pi}{2} n}  (60)^{1/4}\sqrt{\frac{\kend}{k}}
  + {\cal O} \left(\frac{\kend}{k} \right), \quad n=0,1,2,3 \, 
\end{equation}
By accounting for the contribution from the pole closer to the real $t$ axis, and assuming that a single saddle point dominates the integral for $\beta_k$, we find the occupation number for UV modes
\begin{equation} 
\label{eq:UV tail suppression}
n_k^\textsc{(uv)} \simeq 
   \frac{\pi^2}{9}  e^{- 2 \pi \frac{k}{ \kend}} \, .
\end{equation} 
We can see from Fig.~\ref{fig:1_dS-Minkowski_nk_t-k} that the agreement of this approximate formula with numerical results is good for $k>(5-10) \kend$.

This agreement with numerical findings highlights the effectiveness of the steepest descent approach. 
From a computational point of view, the steepest descent approximation is extremely fast and sets no limits on the value of $m_\chi/\HI$ that can be used for the computation. 
On the contrary, the numerical solution for the mode $X_k$ is computational intensive, and the required precision gets quickly unattainable as we increase the particle mass and the exponential suppression on its abundance. 
For these reasons, we rely on the steepest descent approximation to extend the analysis to larger masses when computing the dark matter density and abundance. 
We discuss this in detail in \cref{sec:dmabund}.

\subsection{Power-Law Inflation}
\label{sec:PLI}

As a second scenario, we explore the power-law inflation model~\cite{Lucchin:1984yf, Abbott:1984fp, Martin:2013tda}, which features a scale factor slowly evolving as a power-law, and eventually transitioning to a matter-dominated universe.
Despite being excluded by $\textit{Planck}$ constraints%
\footnote{Power-law inflation can potentially satisfy the current constraints by considering non-canonical scalar fields~\cite{Unnikrishnan:2013vga}.},
this analytical model serves as an excellent test-bed for comprehending the gravitational production of particles with a varying Hubble parameter. 

The motivation to delve into the power-law inflation model is twofold. First, unlike de Sitter spacetime with a constant Hubble rate, the model exhibits a slowly change in the Hubble scale.
This prompts a natural question about the validity of the particle number density formula $\exp(-2\pi \nu)$ as the Hubble rate is evolving: at which time should we evaluate the Hubble rate in this formula? Second, differently from our previous example, the inflationary spacetime evolves into a matter-dominated universe. 
This is the realistic setting after the end inflation when the inflaton oscillates in its minimum. 
We now explore how gravitational particle production proceeds in this cosmological background.

In the inflating phase of power-law inflation, the scale factor scales follows a scaling law given by 
\begin{equation}
a(t) \;  \propto \;  {t}^{\frac{1}{\varepsilon}} \,,
\end{equation}
where $\varepsilon$ is a constant.
The Hubble rate evolves as
\begin{equation}
H(t) = \frac{1}{\varepsilon \, t } \, ,
\end{equation}
and $\varepsilon$ is equal to the slow-roll parameter for the variation rate of the Hubble scale, $\varepsilon \equiv - \frac{ \dot{H}}{H}$.
Power-law inflation then has constant slow-roll parameter $\varepsilon$, and can be realized by an inflation potential with an exponential form, 
\begin{equation}
V(\phi) = M^4 e^{- \sqrt{2 \epsilon} \, \phi / \MP} \, ,
\end{equation}
where $M$ is a dimensionful constant and $\MP$ is the Planck mass. 
However, the inflation potential is somewhat incomplete, as it requires the addition of an exit mechanism to enable the transition to a decelerating phase after inflation. 
To bridge the two phases, we model phenomenologically the scale factor as 
\begin{equation}
a(t) \propto \frac{(t/\tend)^{1/\epsilon} }
   {1 + (t/\tend)^{1/\epsilon} / (t/\tend)^{3/2}}
\label{eq:scalefactorpli}
\end{equation}
This ensures that the scale factor is proportional to $t^{1/\epsilon}$ during the power-law inflation phase ($t \ll \tend$), and smoothly transitions into matter-domination as $t\gg\tend$, with $a\sim t^{3/2}$.

\begin{figure}\centering
Power-law inflation\\
\includegraphics[width=0.5\textwidth]{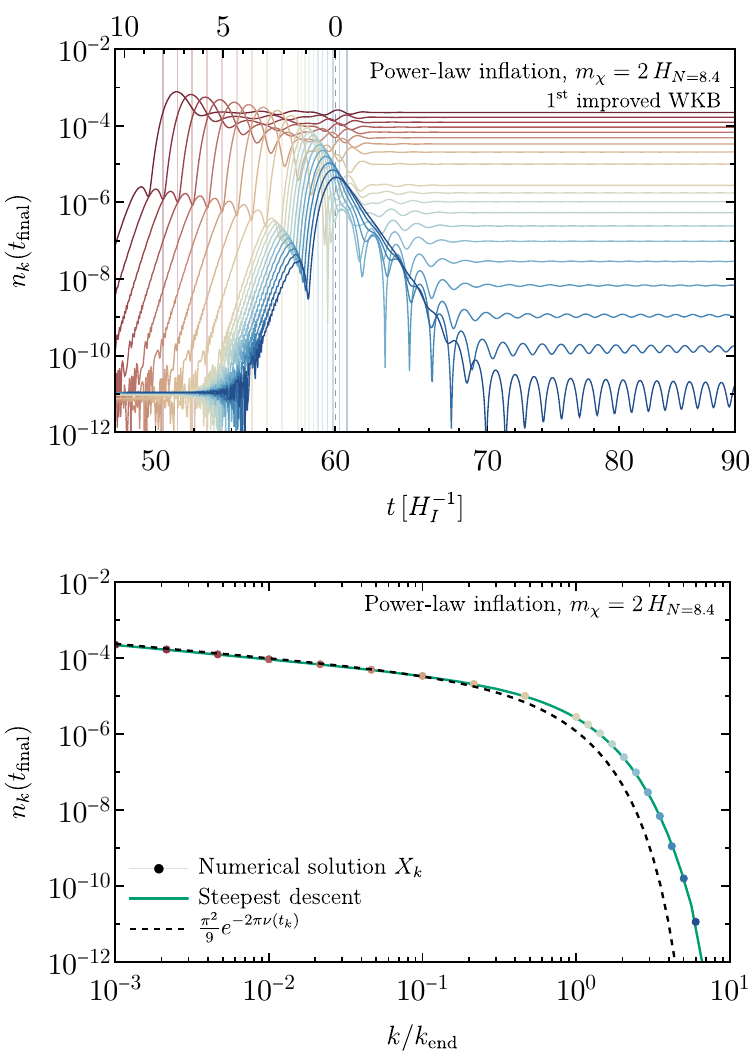}  \hspace{-1em}\hfill
\includegraphics[width=0.5\textwidth]{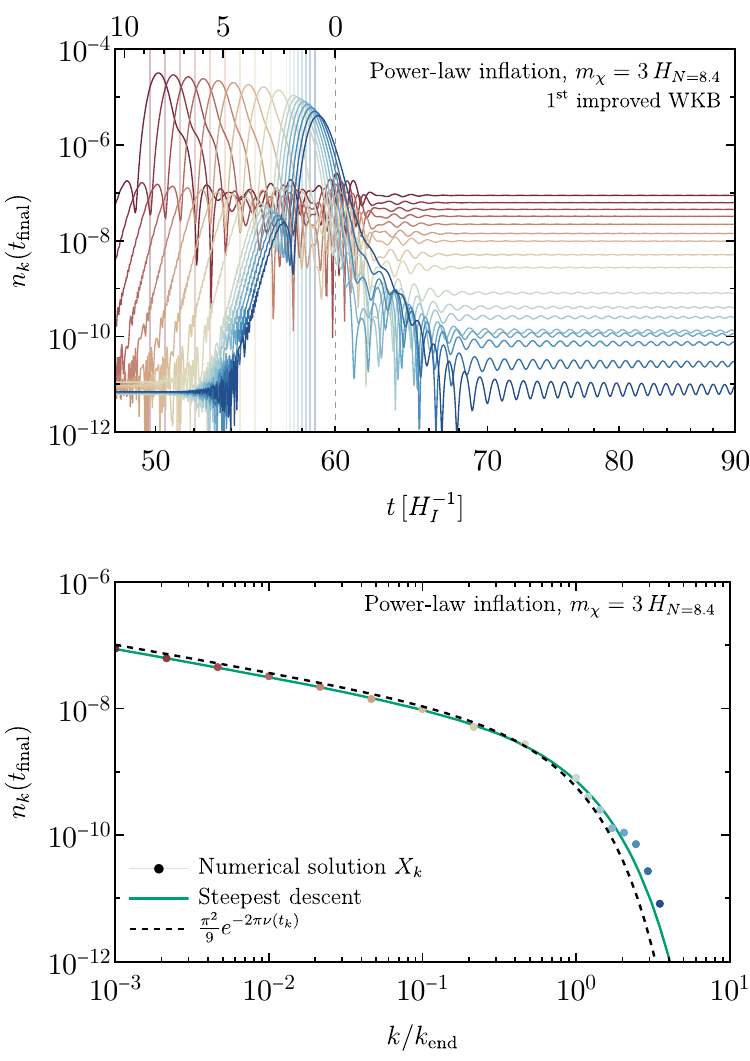}
\caption{Same as Fig.~\ref{fig:1_dS-Minkowski_nk_t-k} for the power-law inflation model.
Here, we fix the units of $m_\chi/\HI$ to the time at 8.4 $e$-folds before the end of inflation. 
For the top panels, the upper horizontal axis marks the number of $e$-folds until the end of inflation, and the vertical colored lines show the value of the time $t_k$ (for the evaluation of $\exp(-2\pi\nu(t_k))$ in the lower panel) for each mode with the corresponding color.
When accounting for the time dependence of the inflationary Hubble rate, the gravitationally produced abundance can be easily estimated via Eq.~\eqref{eq:2pi nu(tk)}, where a constant $\nu$ is replaced by $\nu(t_k)$, with $t_k$ occurring slightly before Hubble crossing.
}
\label{fig:2_Power-Law_nk_t-k}
\end{figure}

We assess the particle occupation number $n_k = |\beta_k|^2$ through numerical solution of $X_k$, and compare it with the steepest descent approximation for $\beta_k$.
The summarized results are presented in \cref{fig:2_Power-Law_nk_t-k} with the choice of parameters $\epsilon= 0.02$, $\tend = 60 \HI$, where we set the arbitrary time normalisation so that $\HI\equiv H_{N=8.4}$ is the Hubble rate at 8.4 $e$-folds before the end of inflation.%
\footnote{The Hubble rate in power-law inflation is $H=\frac{1}{\epsilon t}$, and the cosmic time in our computation has units $H^{-1}$, where $H=1$ at a time $t=1/\epsilon = 50$. At that time, $8.4$ $e$-folds are left to reach $60$ $e$-folds of inflation.}
The mode leaving the Hubble radius at the end of inflation is $\kend \equiv \dot{a}(\tend) = \aend\Hend$.

\Cref{fig:2_Power-Law_nk_t-k} illustrates the occupation number $n_k$ as a function of time $t$ in unit of $\HI$ (top panels) and as a function of the rescaled momentum $k/\kend$ (bottom panels) for two mass choices $m_{\chi} = 2\HI$ (left panels) and $m_{\chi} = 3\HI$ (right panels).
Notably, long-wavelength (IR) modes display a non-flat behavior with a negative slope, while, consistently with our previous observation, short-wavelength (UV) modes exhibit a strong exponential suppression in $k$.
Furthermore, we find consistent results between the full numerical solutions and the steepest descent approximation, that we discuss in the following.

\subsubsection*{Steepest descent method for a time-varying Hubble rate}
As discussed in \cref{sec:steepest}, the first step is to solve $\omega(t_p) = 0$. 
In the limit of small $\epsilon$, and denoting $t_p = |t_p| e^{i \theta_p}$, we find
\begin{equation}
\label{eq:pole PLI}
\omega^2 (t_p) \simeq \frac{k^2}{ a^2(|t_p|)} e^{- 2 i \theta_p / \epsilon} +  m_\chi^2 -\frac{9}{4} H^2(|t_p|) e^{- 2 i \theta_p} \,.
\end{equation}
There are infinite saddle points with a similar real part, and separated by $2\pi i$.
The saddle point $t_k$ lying closer to the real $t$ axis in the lower half-plane, which contributes the most to $\beta_k$, has modulus and argument given by
\begin{equation}
\label{eq:def tk}
\frac{k^2}{ a^2 (|t_k|) } = m_\chi^2 -\frac{9}{4} H^2(|t_k|)  \, , \quad
\theta_k  \simeq -\frac{\pi } {2}\epsilon \,, 
\end{equation}
so that the first term in \cref{eq:pole PLI} has a phase close to $\pi$ and cancels the second term.
For $k\lesssim \kend$, we get $|t_k|\lesssim \tend$ and $\Im(t_k) \simeq -\tfrac \pi 2$, similarly to the analogous result for de Sitter in \cref{eq:dS pole Hcrossing}.
Subsequently, we evaluate the exponent in \cref{eq:saddlepoint1},
\begin{equation} 
-2 i \int_{t_p^*}^{t_p} \d t \, \omega_k ( t)  
= - 2i \nu ( |t_p|)
\int_{\frac{\pi}{2} i}^{ - \frac{\pi}{2} i} 
\d (i \theta/ \epsilon) 
\sqrt{e^{-2i\theta/ \epsilon} + 1}= -2\pi\nu (|t_p|)  \, .
\end{equation}
In conclusion, similarly to the de Sitter case, the steepest descent approximation gives an occupation number for particle production during inflation 
\begin{equation}
\label{eq:2pi nu(tk)}
n_k \simeq \frac{ \pi^2}{9} \exp \Big( -2 \pi \nu (t_k) \Big)  \, , \qquad t_k\,: \ \frac{k^2}{a^2(t_k)} = m_\chi^2 - \frac{9}{4} H(t_k)^2.
\end{equation}
where $\nu(t)$ is evaluated at the time $t_k$ when the physical wavelength $(k/a(t_k))^{-1}$ is a factor $\sim \mathcal O(m_\chi/H)$ smaller than the Hubble radius, so that the mode is still well inside the Hubble radius. 
This analytic result is shown with black dashed lines in the lower panels of \cref{fig:2_Power-Law_nk_t-k}.
For comparison, the green solid lines are computed by finding numerically the pole of $\omega_k(t)$, and by computing numerically $\Omega_k(t)$.
Both the analytical and numerical results using the steepest descent method are consistent for IR modes ($k < \kend$). However, the analytic result \eqref{eq:2pi nu(tk)} cannot be directly extrapolated to the UV modes, as the dispersion relation ($w\neq -1$) and the spacetime background are modified from the ones during inflation.
We still show the simple analytical results (black dashed line) for the UV modes for comparison. 
Importantly, the numerical results (green line), obtained with the full dispersion relation, align well with the numerical solutions of $X_k$.

\subsection{Starobinsky Model of Inflation}
\label{sec:starobinsky}

After exploring particle production in de Sitter and power-law inflation, we now delve into realistic inflationary models, to understand the implication for heavy-particle production. 
In this section, we focus on the Starobinsky Model, though our discussion broadly applies to other inflationary models.

\begin{figure}[h!] \centering
\includegraphics[width=0.48\textwidth]{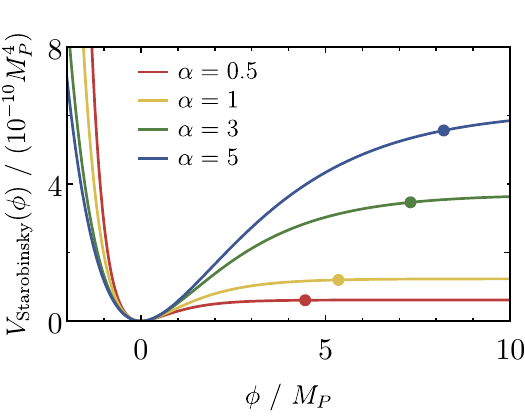}
\hspace{-3em}\hfill
\includegraphics[width=0.48\textwidth]{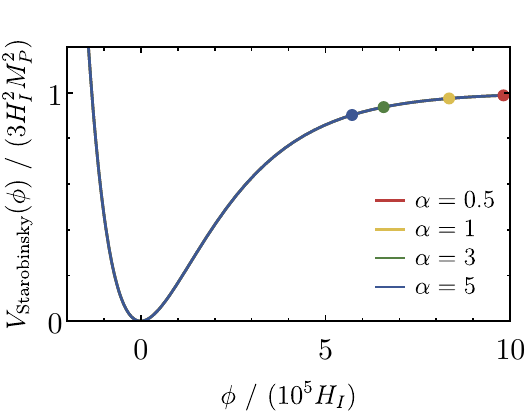}
\vspace{-1em}
\caption{Starobinsky potential for different values of the parameter $\alpha$. 
The dots mark the position of the inflaton field at $N=60$ $e$-folds before the end of inflation.}
\label{fig:Starobinsky_Potential}
\end{figure}

The Starobinsky potential is defined as
\begin{equation}
\label{eq:staropot}
V(\phi) \; = \; \frac{3}{4} \lambda \MP^4 \left(1 - e^{-\sqrt{\frac{2}{3 \alpha}} \frac{\phi}{\MP}} \right)^2 \, ,
\end{equation}
where $\lambda$ and $\alpha$ are real parameters. 
Here, $\lambda$ represents the scale of the inflationary potential, determining the Hubble scale $\HI$ as 
$\HI \simeq \frac{1}{2} \sqrt{\lambda} \MP$. 
The potential is illustrated in \cref{fig:Starobinsky_Potential} in units of $\MP$ (left panel) and $\HI$ (right panel), with $\alpha$ values of $0.5, 1, 3$, and $5$. 

To make contact with CMB observations, we can write down the predicted power spectrum $\cal P_\zeta $, its spectral index $n_s$, and the tensor-to-scalar ratio $r$, 
\begin{equation}
\label{eq:CMBmodel3}
\mathcal{P}_\zeta \simeq \frac{\lambda\, N_*^2}{ 24 \pi^2 \, \alpha } \,, \quad
n_s \simeq 1 - \frac{2}{N_*} \,, \quad r \; \simeq \; \frac{12\alpha   }{N_*^2} \, ,
\end{equation}
where $N_*$ is the number of $e$-folds between Hubble crossing of CMB modes and the end of inflation $\tend$.

The Hubble rate and the inflaton mass at the end of inflation are the crucial parameters governing the particle production rate. 
The Hubble rate slowly evolves over time, starting from a value $\HI \simeq \frac{1}{2} \sqrt{\lambda} \MP$ near the top of the potential.
The inflaton mass after the end of inflation, when the inflation crosses the potential minimum $\phi \simeq 0 $, is given by  
\begin{equation}
m_{\phi} = \sqrt{\frac{\lambda}{\alpha}} \MP  
 \simeq \frac{2}{\sqrt{\alpha}} \HI \, .
\label{eq:mphi_inH}
\end{equation}

Thus, by utilizing CMB observations to determine the scalar power spectrum and spectral index in \cref{eq:CMBmodel3}, 
we fix $\lambda /\alpha$ and therefore the inflaton mass $m_\phi$, while the Hubble rate $\HI\simeq \tfrac 12 \sqrt{\lambda}\MP$ varies with $\lambda$ (as visible from the left panel of Fig.~\ref{fig:Starobinsky_Potential}). 
Alternatively, when using $\HI$ as a reference unit, variations of $\alpha$ correspond to a change of $m_\phi/\HI$.
In the following discussion, we interchangeably refer to variations of $\alpha$ or $m_\phi/\HI$, as encoded in \cref{eq:mphi_inH}).

First, we present the analytic estimation for particle production after inflation. 
Then, on the basis of the analytic results on particle production during and after inflation, we consider two cases: light and heavy inflaton masses, with particle production occurring predominantly during and after inflation, respectively.

\subsubsection*{Analytic estimate of particle production after inflation}
In the previous section, we obtained analytic results for particle production \textit{during} inflation, with the corresponding occupation number given by Eq.~\eqref{eq:2pi nu(tk)}. 
Here, we extend our analysis to estimate, the particle production after inflation, when the inflaton oscillates around the minimum of the potential, with the steepest descent approach. 
This method employs the analytic continuation of the frequency $\omega_k(t)$, which relies on the  analytic extension of the scale factor $a(t)$. 
However, the exact form of $a(t)$ is inaccessible in  the Starobinsky model (or any realistic inflationary model). 
For this reason, our analytic result within the steepest descent approach for particle production after inflation serve as an estimate, yet it remains an essential method to understand the fundamental physics. 
We compute an accurate occupation number through the numerical solution for the mode function $X_k$.

Particle production after inflation is primarily dominated by the first few oscillations of the inflaton $\phi$ (see e.g.~\cite{Brandenberger:2023zpx} for the corresponding production of light fields). 
When oscillating around the minimum, its energy density redshifts as a non-relativistic species, so the envelope of the oscillating $\phi$ goes as $a^{-3/2}\sim 1/t$. 
Because of this redshifting amplitude, the total inflaton energy density (and hence $H(t)$) does not scale smoothly with time, and it features wiggles (although it is monotonically decreasing) of relative size $\sim H^2(t)/m_\phi^2$. 
An alternative description is that the equation of state $w(t)$ oscillates between the values for kination and vacuum energy, accordingly with the inflaton oscillations.
This oscillating pressure induces wiggles in the frequency $\omega_k(t)$ of the superheavy field $\chi$, as shown in \cref{fig:omega_zeroes}.

For the sake of providing a simple analytical formula capturing the main parametric scaling of particle production in this epoch, we use the steepest descent approximation with a simplified parametrisation of the inflaton by neglecting its redshift,
\begin{equation}
   \phi(t) \simeq  - \phi_0 \cos ( m_\phi t ) \, ,
\end{equation}
where $t=0$ marks the first time the inflaton crosses the potential minimum. 
For this simple estimate, we neglect the expansion of the universe within one period of the oscillation, but we take it into account for our numerical results. 
In the present derivation, this approximation is motivated because the first oscillation contributes most significantly to particle production due to a large value of $\phi_0$, and thus of $\sim H^2(t)/m_\phi^2$. 
We can then write the squared frequency of the heavy 
scalar $\chi$ for modes $k<\kend$ as
\begin{equation}
\omega_k^2 (t) \simeq m_\chi^2 + \frac{9}{4} w H^2
   \simeq m_\chi^2 + \frac{3(\dot \phi^2 - m_\phi^2 \phi^2) }{8 \MP^2}
   \simeq m_\chi^2 - \frac{9}{8}\Hend^2\cos(2 m_\phi t)\,,
\end{equation}
where in the last step we used $3\Hend^2\MP^2 \simeq m_\phi^2 \phi_0^2$.
By analytically extending $\omega_k$ to complex $t$, we find the closest saddle point to the real axis as
\begin{equation} 
t_p  \simeq  0 + i \frac {1}{2 m_\phi} \log(\frac{16}{9}\frac{m_\chi^2}{\Hend^2})\, .  
\end{equation} 
The real part of $t_p$ corresponds to when the inflaton crosses its potential minimum, which marks with the peak of particle production.
We can then estimate the particle occupation number produced after inflation with the steepest descent approximation,
\begin{equation}
   n_k \simeq 
      \frac{\pi^2}{9} \exp ( -4 \int_0^{t_p}  {\rm d } t \, \omega_k (t) )  
      \simeq 
      \frac{\pi^2}{9} 
      \exp ( - \frac{4 m_\chi}{m_\phi } \log( \frac{16}{9}\frac{m_\chi^2}{\Hend^2})  ) \, .
\label{eq:nk_reheating}
\end{equation}
This simple and instructive result shows that the gravitational production of the supermassive field $\chi$ grows with the mass of the inflaton field.
However, it is essential to note that this approach breaks down when $m_\phi \gg m_\chi$. 
In that case, as shown in \cref{fig:poles}, the saddle points approach each other closely, making the approximation of a single saddle point in \cref{eq:nk_reheating} invalid. 
The scenario of gravitational production for $m_\phi \gg m_\chi$ is discussed in \cite{Ema:2018ucl,Chung:2018ayg}.

\subsubsection*{Numerical results}
Depending on the inflaton mass $m_\phi$, gravitational production of a heavy particle after inflation (when the inflaton oscillates around its minimum) can overcome the inflationary production.
When $m_\phi \lesssim \HI$ ($\alpha \gtrsim 1$), the later contribution after inflation does not overshoot the inflationary production at Hubble crossing for IR modes $k<\kend$, and their final occupation number can be estimated as $n_k \sim \exp(-2 \pi \nu (t_k))$.
Conversely, if the inflaton mass is heavier than the Hubble scale ($\alpha \lesssim 1$), the oscillation of the inflaton at the end of inflation can lead to the production of heavy particles through gravitational preheating.
In this scenario, the occupation number is approximately $n_k \sim \exp ( - {\cal O}(1) m_\chi / m_\phi )$ as in \cref{eq:nk_reheating}.
Note that the steepest descent method, used to evaluate the gravitational particle production during reheating period, 
$n_k \sim \exp ( - {\cal O}(1) m_\chi / m_\phi )$, 
is not applicable when $m_\phi \gg m_\chi$, as previously explained. 
Given these distinct production mechanisms, we explore 
both regimes by presenting analytic and numerical results for each of them.

\paragraph{Light inflaton mass ($m_\phi \lesssim \HI$ or $\alpha \gtrsim 1$)}

\begin{figure}\centering
Starobinsky model, light inflaton mass: $m_\phi < \HI < m_\chi$
\includegraphics[width=0.5\textwidth]{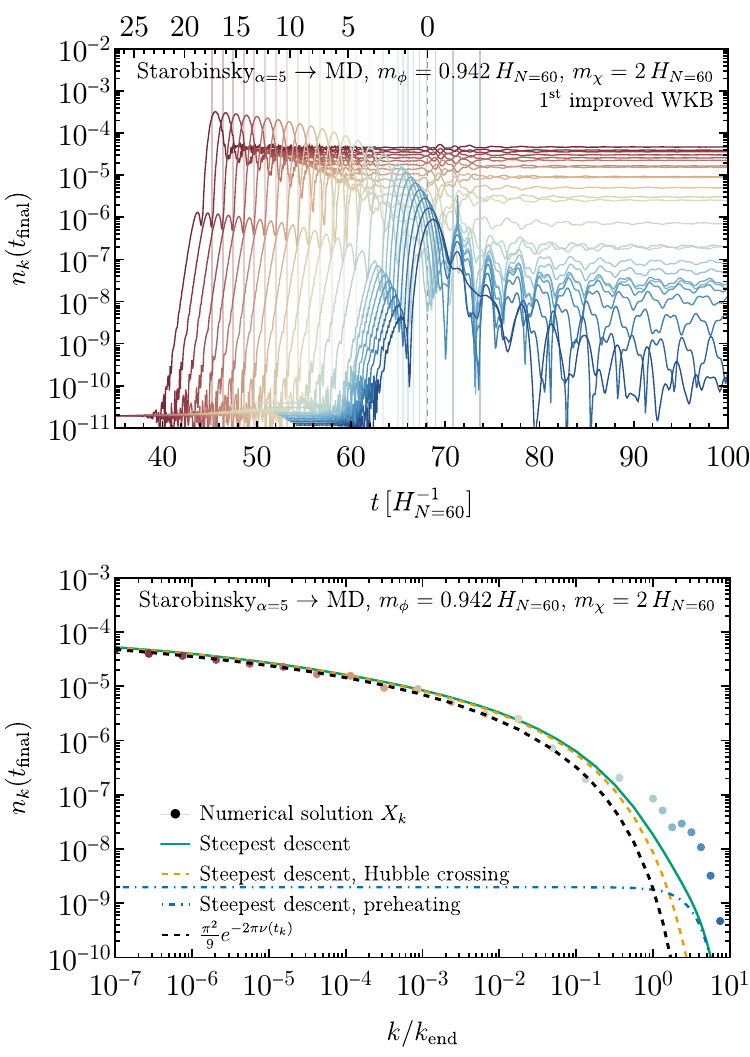} \hspace{-1em}\hfill
\includegraphics[width=0.5\textwidth]{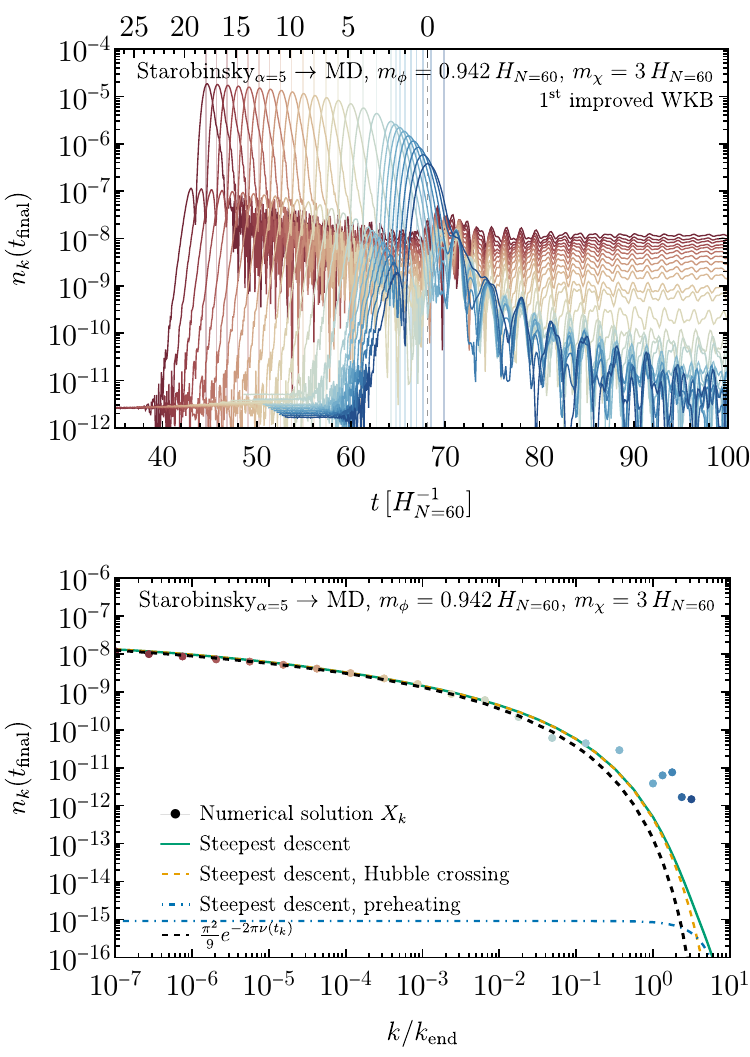} \hspace{-1em}\hfill
\caption{
Same as Fig.~\ref{fig:2_Power-Law_nk_t-k}, for the Starobinsky inflationary model with $\alpha=5$.
Here, we define $\HI \equiv H_{N=60}$ at 60 $e$-folds before the end of inflation. 
In the bottom panels, we show separately the production during inflation (`Hubble crossing', dashed yellow) and right after the end of inflation (`preheating', dot-dashed blue), along with their sum (solid green), estimated with the steepest descent approximation.}
\label{fig:3_Starobinsky_a5_nk_t-k}
\end{figure}

We consider a light inflaton with $m_\phi \lesssim \HI < m_\chi$, resulting in suppressed particle production 
post-inflation. Here, the Hubble scale is defined at $60$ $e$-folding before the end of inflation, $\HI = H_{N=60}$. 
In accordance to \cref{eq:mphi_inH}, a lighter inflaton mass corresponds to a larger value of $\alpha$.
As a benchmark model, we select $\alpha = 5$, which yields an inflaton mass $m_\phi$ (in its minimum at the end of inflation) smaller than the inflationary Hubble, $m_\phi = 0.94 \HI$. 
We choose this value where the post-inflationary gravitational production has a very moderate impact on the spectrum of $\chi$, in order to highlight the difference between the two regimes.
We compare the numerical evaluation of the mode function $X_k$, with the analytic formula for particle production during inflation with a time-varying Hubble rate, 
expressed in \cref{eq:2pi nu(tk)}, as well as with the numerical evaluation of the steepest descent method.

In the first numerical approach, we solve the evolution of the modes $X_k$ within the Starobinsky background, and evaluate the occupation number using the 1st order improved WKB approximation. 
The results are presented in \cref{fig:3_Starobinsky_a5_nk_t-k} with colored lines for each mode $k$ for two scalar masses $m_\chi = 2, 3 \HI$.
The upper plots show $n_k$ as a function of cosmic time $t$ for each mode, and the bottom plots show the corresponding value of $n_k$ (with the same color code) at a late time $t=200\,\HI$ (with a time average over the small decreasing oscillations).

In the lower plots, the spectrum of the IR modes ($k \lesssim 0.1 \kend$, in red-orange) align very well with the analytical estimate of inflationary particle production with the steepest descent method,
$n_k \simeq \frac{\pi^2}{9} \exp(-2\pi \nu(t_k))$ (black dashed line), as discussed in \cref{sec:PLI}.
We also estimate $n_k$ by numerically finding the dominant poles during inflation and shortly after $\tend$, and performing the integral of \cref{eq:saddlepoint1}. 
In order to do so, it is necessary to analytically extend $\omega_k(t)$. 
For the pole around Hubble crossing (see \cref{fig:poles}), one can easily fit the (slowly varying) Hubble rate with a polynomial in $t$, and compute the scale factor $a(t)\sim \exp(\int H(t))$.
The analytical extension of $\omega_k(t)$ around the inflaton oscillations (blue band in \ref{fig:omega_zeroes}) is more delicate: we perform a fit of $\phi(t)\sim \tfrac{1}{t-t_0} \cos(\omega_\phi t)$ around a few inflaton oscillations to take care of the term $\tfrac 94 w H^2$ in $\omega_k(t)$, and fit $a\sim t^{2/3}$ for the gradient term $k^2/a^2(t)$.
This numerical method yields consistent results in both the IR and UV region.
The sum of the contributions from the two poles, as in \cref{eq:saddlepoint1}, is depicted as a solid green line in the lower plots. 
For UV modes ($k>\kend$, marked as blue dots), particle production during inflation is exponentially suppressed in $k$, so that post-inflationary production becomes the dominant contribution.
Finally, we notice that the occupation number derived through the steepest descent method underestimates the one obtained through the numerical solution of $X_k$. 
This may be due to the inaccuracy of the analytical extension of $\omega_k(t)$ in the complex $t$ plane, which can hardly be fitted over a large enough time range on the real axis.

\paragraph{Heavy inflaton mass ($m_\phi \gtrsim \HI$ or $\alpha \lesssim 1$)}
\begin{figure}\centering
Starobinsky model, heavy inflaton mass: $\HI<m_\phi < m_\chi$
\includegraphics[width=0.5\textwidth]{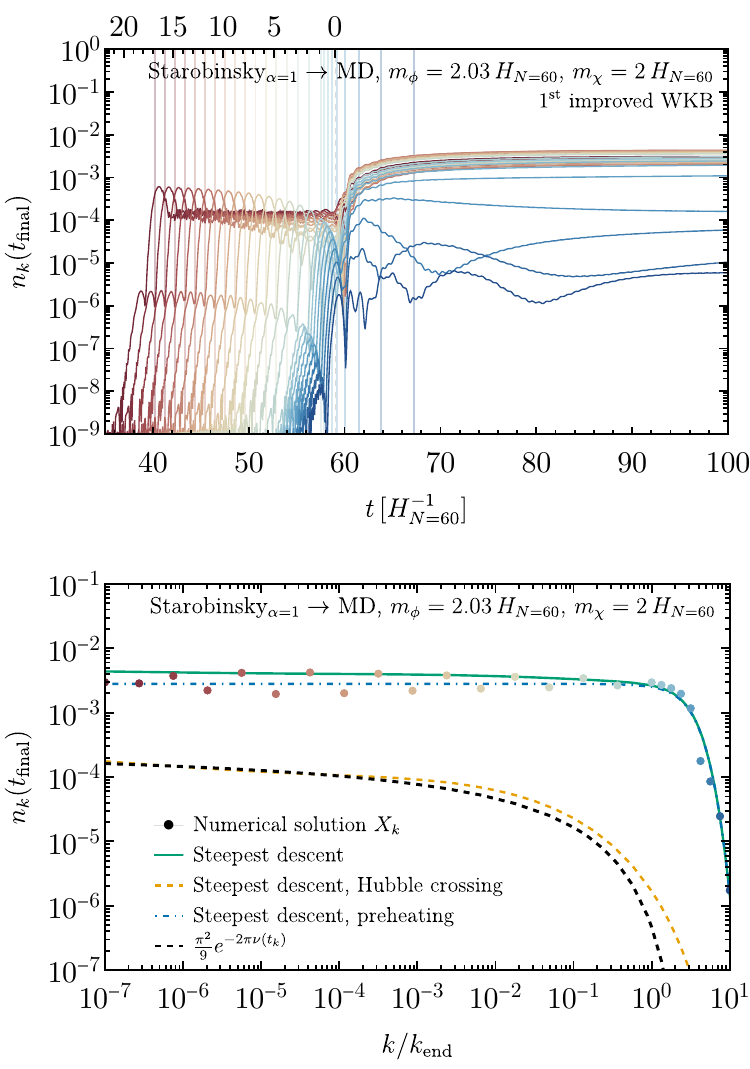} \hspace{-1em}\hfill
\includegraphics[width=0.5\textwidth]{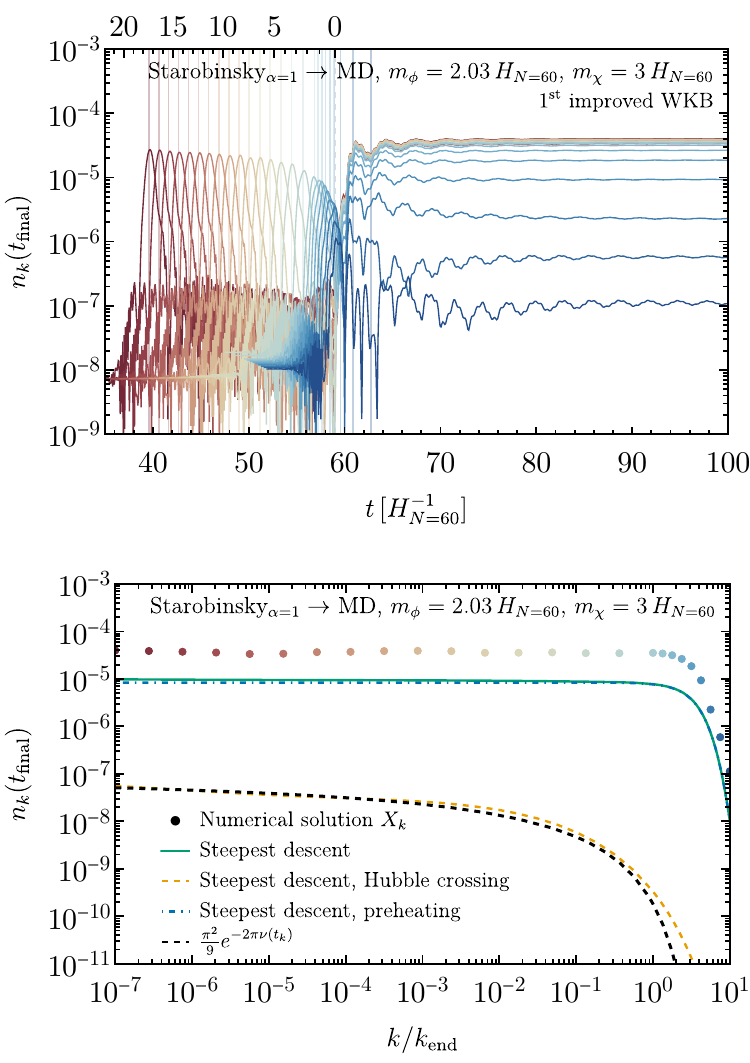} \hspace{-1em}\hfill
\caption{Same as in Fig.~\ref{fig:3_Starobinsky_a5_nk_t-k}, for the case of heavier inflaton $m_\phi =2.03\HI$ ($\alpha=1$) in the Starobinsky model.
When the inflaton mass $m_\phi$ in its minimum is larger that the inflationary Hubble rate (but smaller than $m_\chi$, post-inflationary gravitational production can overcome the inflationary production, with an occupation number in the IR mode (red-orange lines and dots) $n_k\sim \exp(-\mathcal O(1)m_\phi/m_\chi)$.
The dot-dashed blue lines in the bottom plots highlight the impact of post-inflationary gravitational production with respect to the inflationary one (dashed yellow line).}
\label{fig:3_Starobinsky_a1_nk_t-k}
\end{figure}

We focus now the case where the inflaton mass in its minimum is large ($m_\phi \gtrsim \HI$), but $m_\phi$ is not significantly larger than the spectator field mass $m_\chi$.
Under these conditions, particle production primarily occurs \textit{after} inflation for both UV and IR modes.
We adopt $\alpha = 1$ as our benchmark model, which corresponds to an inflaton mass $m_\phi = 2.03 \HI$. 
The numerical solutions, along with the results obtained with the steepest descent method, are summarized in \cref{fig:3_Starobinsky_a1_nk_t-k}.
We show the results for two scalar masses: $m_\chi = 2 \HI$ (left column) and $m_\chi = 3 \HI$ (right column). 
The upper panels of the figure depict the rise in $n_k$ immediately after the end of inflation ($\tend \approx 59\,\HI$, corresponding to 0 $e$-folds until the end of inflation on the upper axis), indicating that the predominant particle production occurs after inflation, during the preheating phase.
This observation is corroborated in the lower panel by comparing the numerical solution (red/blue colored points) with the lines obtained from the steepest descent method. 
The dotted black line (given by $n_k \simeq \frac{\pi^2}{9} \exp(-2 \pi\nu(t_k))$) is consistent with the numerical evaluation of the saddle point approximation using only the saddle during inflation (dashed yellow line).
The two lines are in agreement in the IR spectrum, while they show some deviation in the UV spectrum.
Moreover, post-inflation particle production, as determined by the steepest descent method, is several orders of magnitude greater than inflation production.
For $m_\chi = 2 \HI$, numerical solutions for $X_k$ (red-blue points) agree with the steepest descent method (green line), whereas for $m_\chi = 3 \HI$, the prediction from the steepest descent method is slightly smaller. 
This discrepancy for large $m_\chi$ is expected, because it implies a wider extrapolation from the value $\omega_k\sim m_\chi$ on the real $t$ axis (see Fig.~\ref{fig:omega_zeroes}) to $\omega_k(t_n)=0$ in the complex plane (see Fig.~\ref{fig:poles}). 
This results in an amplification of the error of the analytic continuation of $\omega_k(t)$.
\begin{figure}\centering
\begin{tikzpicture}
\hspace{-.12\textwidth}
\node [above right,inner sep=0] (image) at (0,0) {
\includegraphics[width=1.2\textwidth]{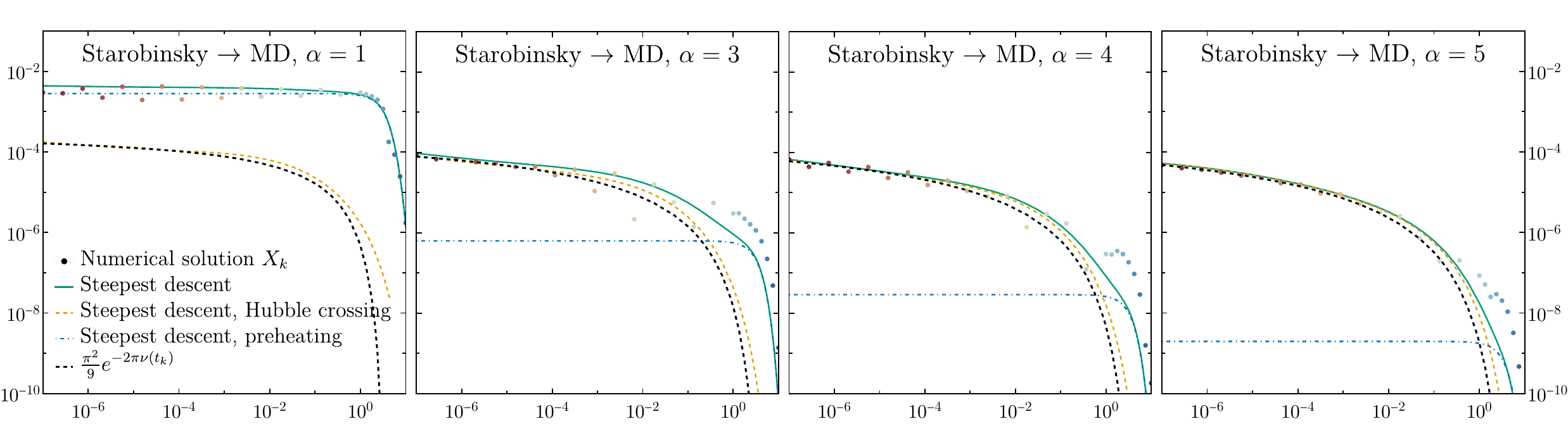} 
};
\begin{scope}[
x={($0.05*(image.south east)$)},
y={($0.05*(image.north west)$)}]
\node at (10,20) {Occupation numbers of a superheavy $\chi$ in the Starobinsky model for decreasing $m_\phi/\HI$};
\node at (10,-1) {$k/\kend$};
\node[label={[label distance=0mm,text depth=-1ex,rotate=90]above:$n_k(\tfinal)$}] at (0,10) {};
\end{scope} 
\end{tikzpicture}
\caption{Comparison of the particle production for decreasing values of the inflaton mass in units of $\HI$ (respectively $m_\phi =2.0,\,1.2,\,1.0,\,0.9\,\HI$, corresponding to increasing $\alpha=1,\,3,\,4,\,5$ in the Starobinsky model), with the same color code as the bottom panels of Fig.~\ref{fig:3_Starobinsky_a5_nk_t-k} and \ref{fig:3_Starobinsky_a1_nk_t-k}. 
For all plots, the spectator field mass is $m_\chi=2 H_{N=60}$.
The inflationary production (dashed yellow and black lines) stays the same in the four cases, while the post-inflationary production decreases, as shown by the red-blue points (numerical solution of the modes $X_k$) and the solid green line (steepest descent approximation).
}
\label{fig:3_Starobinsky_nk}
\end{figure}
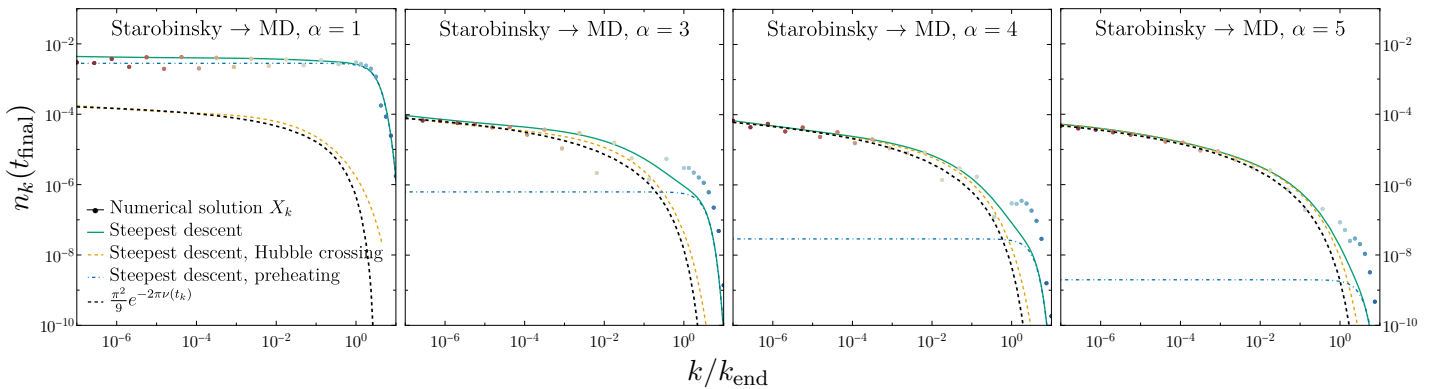

Finally, to illustrate the influence of inflaton mass, and thus the model dependence, on the post-inflationary gravitational production, we plot the occupation number $n_k$ 
against momentum $k$ for four different values of $m_\phi/\HI$ in \cref{fig:3_Starobinsky_nk} (with the same spectator mass $m_\chi = 2 \HI$).
The inflaton mass dependence is represented by various values of $\alpha$ ($1, 3, 4$, and $5$) in the four panels. 
The dashed blue lines show the increase with $m_\phi$ (or decrease with $\alpha$) of the post-inflationary particle production.

\section{Reheating and Dark Matter Abundance}
\label{sec:dmabund}
The goal of this section is to compute the present abundance of the gravitationally produced $\chi$, in order to assess the parameter space where it can form the totality of dark matter.
We first discuss the reheating epoch that occurs immediately after the end of inflation in a matter-dominated universe. 
After $\tend$, as the inflaton decays, the thermal plasma dilutes until it reaches a maximum temperature $T_{\rm max}$~\cite{Giudice:2000ex}. 
Subsequently, the temperature decreases 
until reheating, which occurs when the energy density of the inflaton becomes equal to the energy density of radiation, $\rho_{\phi}(a_\rh) = \rho_{R}(a_\rh)$.
The reheating temperature can be expressed as a function of the inflaton decay rate $\Gamma_{\phi}$ as~\cite{Garcia:2020eof, Pallis:2005bb}
\begin{equation}
\label{eq:reheatingfit}
\frac{\pi^2 g_{*,\rh} T_{\rh }^4}{30} \; \simeq \; 3  \left(\frac 25\Gamma_{\phi} \MP\right)^2 
  \ \Rightarrow \
  T_{\rh} \simeq 1.7 \times 10^{15} \, \GeV  \cdot \,  g_{*,\rh}^{-1/4} \cdot \left(\frac{\Gamma_{\phi}}{10^{12} \, \rm{GeV}} \right)^{1/2} \,,
\end{equation}
where $g_{*,\rh}$ is the total number of relativistic degrees of freedom of the thermal bath at $T_\rh$.

For the case of superheavy dark matter that we consider in this paper, the number-density $\tfrac{1}{2\pi^2}k^3 n_k$ is blue-tilted, and the abundance is dominated by the modes around the Hubble radius towards the end of inflation, with $k/\kend \sim \mathcal O(1 - 10)$. 
The shorter-wavelength UV tail, with $k/\kend > 10$, is exponentially suppressed, as shown by \cref{eq:UV tail suppression}.
We can evaluate the particle occupation number $\Nk = |\beta_k|^2$ at some time after the end of inflation, when the comoving number density $\nco$ is constant:
\begin{equation}
\label{eq:nco}
\n(t) \, a(t)^3  =
  \frac{1}{2\pi^2} \int_0^{\infty} dk \, k^2 |\beta_k|^2 \equiv \nco  = \text{const}. ,
\end{equation}
where $\n(t)$ is the physical number density and $a(t)$ is the scale factor at a time $t> \tend$ when the particle number density remains constant. 
We evaluate the comoving particle number density numerically to compute the present-day dark matter abundance.  

The reheating phase is complete when the energy density of the inflaton $\rho_\phi$ becomes equal to the energy density of radiation $\rho_R$:
\begin{equation}
\label{eq:rehpars}
\rho_R(a_\rh) =  \rho_{\phi}(a_\rh) 
 = \rho_{\phi}(\aend) \left(\frac{\aend}{a_\rh}\right)^3 
 = 3\MP^2 \Hend^2 \left(\frac{\aend}{a_\rh}\right)^3 \, .
\end{equation}
After $a_\rh$, the universe becomes radiation-dominated and the dark matter yield remains constant:
\begin{equation}
Y_{\chi}(a_\rh) \equiv \frac{n_{\chi}(a_\rh)}{s(a_\rh)} \; = \; \frac{\nco \aend^{-3} \,T_{\rh}}{4 \MP^2 \Hend^2} \, ,
\end{equation}
where we introduced the entropy density $s(T) = \tfrac{2 \pi^2}{45} g_* T^3$ and we used \cref{eq:nco} and 
\cref{eq:rehpars}.
The present relic abundance of dark matter is given by
\begin{equation}
\Omega_{\chi} h^2 = \frac{\rho_{\chi}(a_0)}{\rho_\text{cr}} 
  = \frac{m_{\chi} n_{\chi}(a_0)}{\rho_\text{cr}} h^2
  = \frac{m_{\chi} Y_\chi(a_{\rh}) s_0}{\rho_\text{cr}} h^2 
  \, ,
\end{equation}
where $a_0$ is the scale factor today,
$\rho_\text{cr} = 1.054 \times 10^{-5} \, h^2 \,\GeV\,\rm{cm}^{-3}$ is the critical density today, $h=H_0/(100\text{km/s/Mpc})$ and $s_0 = 2891.2 \, \rm{cm}^{-3}$ is the present entropy density.
Using \cref{eq:reheatingfit}, we finally obtain
\begin{equation}
\label{eq:dmabundance1}
\begin{aligned}
\Omega_{\chi} h^2 
 & = 6.9 \cdot 10^7 
   \parfrac{m_\chi}{\GeV}
   \parfrac{\nco \aend^{-3} T_\rh}{\MP^2 \Hend^2} = \\
 & = 0.12\,\parfrac{g_{*,\rh}}{106.75}^{-1/4} 
   \frac{\nco}{6\cdot 10^{-10}\,\kend^3}
   \frac{m_\chi}{\Hend} 
   \parfrac{\Gamma_\phi}{10^{-3}\Hend}^{1/2}
   \parfrac{\Hend}{10^{12} \, \GeV}^{5/2}\, .
\end{aligned}
\end{equation}
\begin{figure}[h!]\centering
\includegraphics[width=0.5\textwidth]{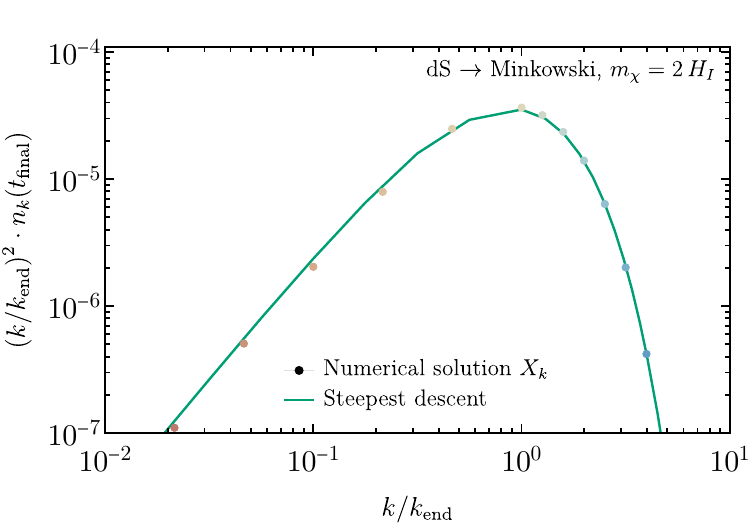} \hspace{-1em}\hfill
\includegraphics[width=0.5\textwidth]{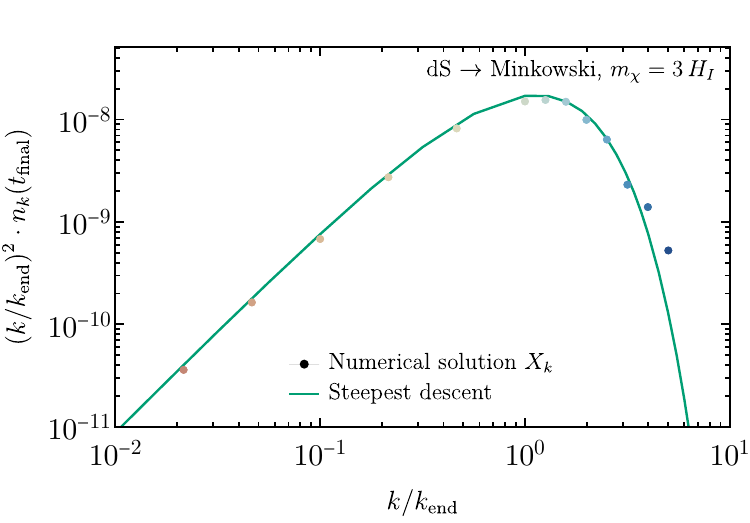}
\includegraphics[width=0.5\textwidth]{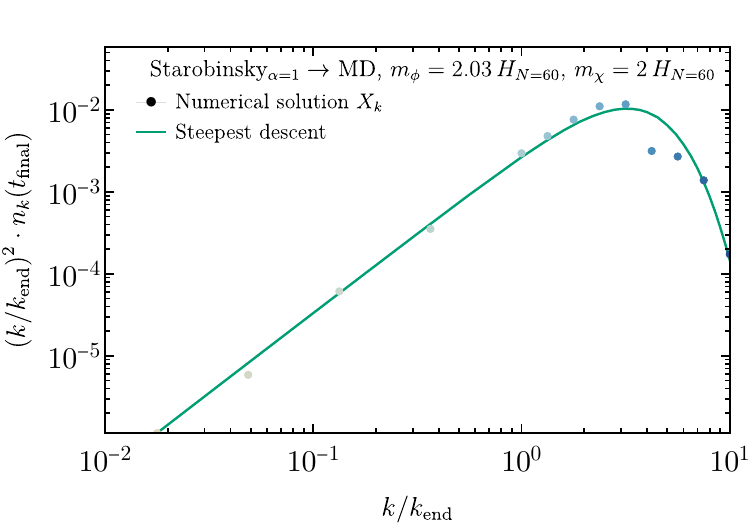} \hspace{-1em}\hfill
\includegraphics[width=0.5\textwidth]{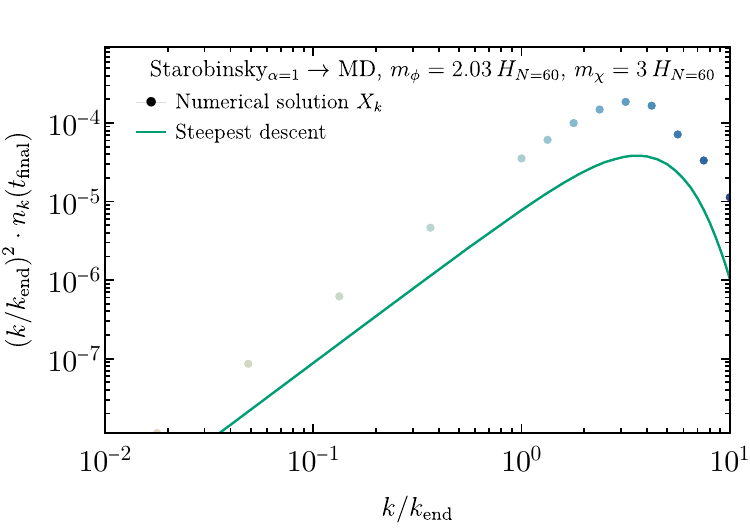}
\caption{The phase space distribution $(k/\kend)^2 \cdot n_k(\tfinal )$ of the integral~\eqref{eq:nco} as a function of $k/\kend$ for the pure de Sitter model (top panels) and the Starobinsky model with post-inflationary production (bottom panels). 
The solid green line shows the steepest descent approximation.}
\label{fig:k2nk}
\end{figure}

We are now able to compute the present abundance for the gravitationally produced dark matter in two of the models considered in this paper.
We illustrate the integrand of the number density in \cref{eq:nco} in Fig.~\ref{fig:k2nk} for the inflationary models of de Sitter (\cref{sec:dstomink}) and Starobinsky (\cref{sec:starobinsky}).
In the top panels, we show (for two dark matter masses $m_{\chi} = 2\HI$ and $3\HI$) the numerical solution (red/blue dots) of the model dS $\to$ Minkowski, together with the steepest descent approximation (green line). 
The two methods are in excellent agreement. 
The slope of the integrand scales as $(k/\kend)^2 \cdot n_k(\tfinal ) \propto k^2$ for the IR modes $k \ll \kend$, while the UV tail $k \gg \kend$ is exponentially suppressed.
The bottom panels show the same results for the model of Starobinsky inflation with $\alpha=1$ (case of heavy inflaton mass, with significant post-inflationary production. 
As discussed at the end of \cref{sec:starobinsky}, the green line for the steepest descent method offers a worse approximation of the result at larger $m_\chi$, because of the bigger extrapolation that is implied in the analytical extension of $\omega_k(t)$.

\begin{figure}[h!]\centering
\vspace{-10pt}
\includegraphics[width=0.6\textwidth]{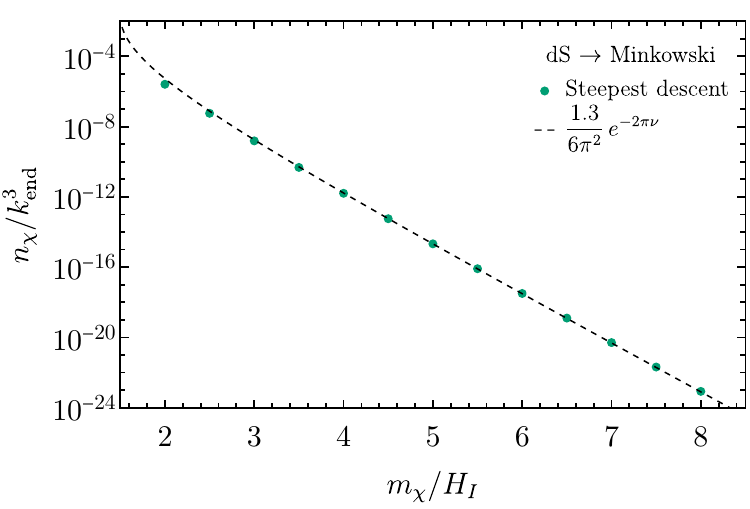}
\vspace{-10pt}
\caption{Total comoving number density $\nco$ of superheavy particles in the de Sitter inflationary model, and fit with the formula in Eq.~\cref{eq:nco dS}.}
\label{fig:ntot_dS}
\end{figure}

The total comoving number density of dark matter particles in the de Sitter model can be fitted by
\begin{equation}
\label{eq:nco dS}
\nco
\simeq \frac{1.3}{6\pi^2}  \kend^3 e^{-2 \pi \nu} \quad \text{(de Sitter inflation)}  \, ,
\end{equation}
as shown in Fig.~\ref{fig:ntot_dS} together with the points computed with the saddle point approximation.%
\newline
The corresponding dark matter abundance from \cref{eq:dmabundance1} is
\begin{equation}
\label{eq:OmegaDM dS}
\Omega_{\chi} h^2 \; \simeq \; 1.5 \times 10^6\, e^{-2 \pi \nu} \left(\frac{m_{\chi}}{\rm{GeV}} \right) \left(\frac{\Hend T_{\rh}}{\MP^2} \right) \quad \text{(de Sitter inflation)} \, .
\end{equation}
Imposing the observed dark matter constraint $\Omega_{\chi} h^2 \simeq 0.12$, we obtain the relation between $\Hend,\,T_\rh,\, m_\chi$ that gives the correct dark matter abundance:
\begin{equation}
\label{eq:DM TRH dS}
T_{\rh} \simeq 
  4.7 \cdot 10^{14}\GeV \cdot \frac{10^{-9}}{e^{-2\pi \nu}} \cdot \frac{(10^{12}\GeV)^2}{m_{\chi} \Hend} 
\quad \text{(de Sitter inflation)} \, .
\end{equation}
The reheating temperature $T_\rh$ can be traded for the inflaton decay rate $\Gamma_\phi$ through \cref{eq:reheatingfit}. 
We can further relate $\Gamma_\phi$ to the Hubble rate $\Hend$, because $m_\phi\sim \Hend$, and the decay rate $\Gamma_\phi$ can be related via dimensional arguments to  $m_\phi\sim \Hend$. We can then introduce the ratio $\Gamma_\phi/\Hend$, which quantifies the strength of preheating and the size of the inflaton couplings (for a perturbative decay rate controlled by a coupling $y$, and taking $m_\phi \sim \Hend$, we would have $\Gamma_\phi/\Hend \sim y^2/(8\pi)$).
This leaves us with two free parameters: $m_\chi$, and $T_\rh$ or $\Hend$.
\begin{figure}[h!]\centering
Dark matter parameter space for de Sitter inflation \vspace{-1em}\\
\includegraphics[width=0.49\textwidth]{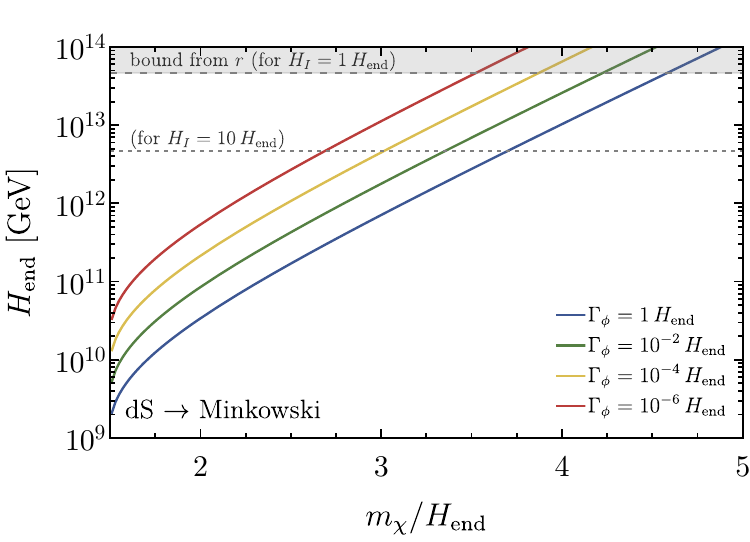}\hspace{-1em \hfill}
\includegraphics[width=0.49\textwidth]{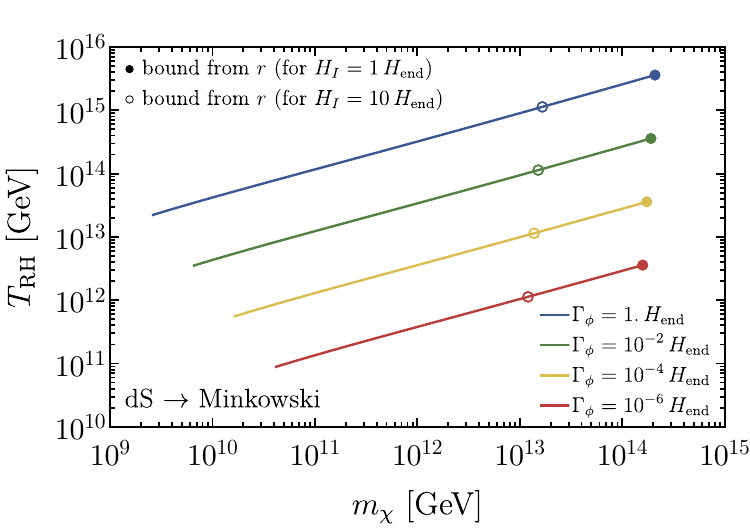}
\vspace{-1em}
\caption{Parameter space for gravitationally produced superheavy dark matter in the de Sitter inflationary model.
The observed relic abundance is achieved when \cref{eq:DM TRH dS} is satisfied, and the parameters $T_\rh$ and $\Hend$ can be related through the inflaton decay rate $\Gamma_\phi$. 
We trade one of the two parameters ($T_\rh$ on the \textit{left} plot, $\Hend$ on the \textit{right}) for different assumptions on the strength of reheating $\Gamma_\phi/\Hend$.
We also show with dashed horizontal lines (\textit{left}) and dots (\textit{right}) the upper limits on the inflationary $\HI$, and on $T_\rh$ on the right plot, from Planck/\textsc{bicep}, assuming $\HI =1,10\,\Hend$.
In both plots, the end-points of the colored curves at low $m_\chi$ mark the regime of massive $m_\chi>\tfrac 32\HI$ that we assume in our analysis.}
\label{fig:DM_dS}
\end{figure}
We show in Fig.~\ref{fig:DM_dS} the dark matter abundance for the de Sitter inflationary model for these two choices. 
The left panel refers to the inflationary quantities $m_\chi/\Hend$ and $\Hend$, and shows the dark matter abundance for different choices of $\Gamma_\phi/\Hend$ and two values of $\HI/\Hend$.
The corresponding upper limit on $\HI$ of Planck and \textsc{bicep} from the non-detection of primordial B-modes in the CMB polarisation ($r<0.036$ at 95\% CL \cite{BICEP:2021xfz}) is shown with horizontal dashed lines.
The right panel shows the same curves in the parameter space $(m_\chi,T_\rh)$, with the dots marking the upper limits on $\HI$.
In both plots, the limiting points on the left correspond to $m_\chi = 3\HI/2$, below which our analysis for superheavy spectator fields does not apply.

\begin{figure}[h!]\centering%
\vspace{-10pt}%
\includegraphics[width=0.6\textwidth]{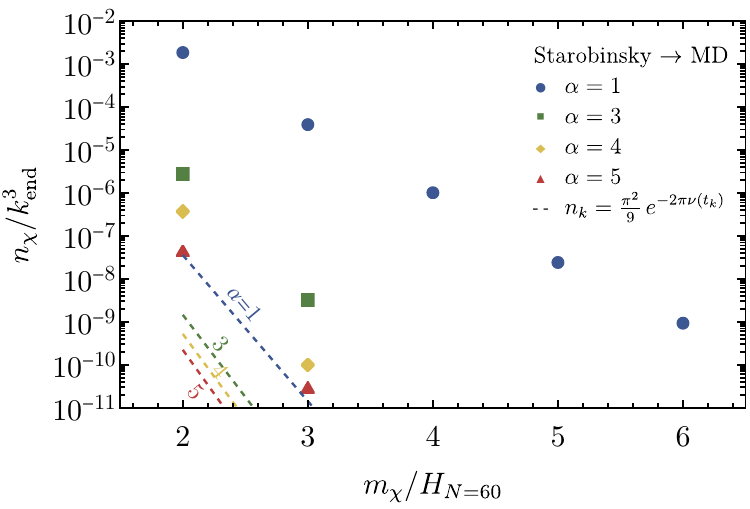}%
\vspace{-10pt}
\caption{Total comoving number density $\nco$ of superheavy particles in the Starobinsky inflationary mode, computed with the numerical results of the evolution of the mode functions $X_k$. 
Decreasing values of $\alpha$ (marked with different colors and point markers) correspond to increasing the inflaton mass in its minimum in Hubble units, and thus enhancing the post-inflationary production. 
The dashed lines show for each $\alpha$ (with the same color code) the predicted abundance if the post-inflationary production is neglected. 
The dots for $\alpha=1$ are well fitted by \cref{eq:nco Starobinsky}.}%
\label{fig:ntot_Starobinsky}%
\end{figure}

For the Starobinsky model of inflation discussed in \cref{sec:starobinsky}, we compute the total number density from the numerical results obtained by evolving the mode equations for $X_k$ (shown in Figs.~\ref{fig:3_Starobinsky_a1_nk_t-k} and \ref{fig:k2nk}). 
The results is shown in Fig.~\ref{fig:ntot_Starobinsky}, and the case $\alpha=1$ is well fitted by
\begin{equation}
\label{eq:nco Starobinsky}
\nco 
  \simeq \frac{76}{6\pi^2} \kend^3 e^{-0.55 \pi \nu} 
  \quad \text{(Starobinsky inflation, $\alpha=1$)} \, .
\end{equation}

The exponential suppression featured in realistic inflationary models is milder than the standard $\exp(-2\pi\nu)$ for inflationary production at Hubble crossing, as a result of the enhanced production after the end of inflation (especially for larger inflaton masses). 
The UV tail $k\gtrsim \kend$ dominates this result, as visible in \cref{fig:k2nk}, and increases by a factor $\sim \mathcal O(1)$ the predicted mass for dark matter. 
The relic abundance obtained from \cref{eq:dmabundance1} is 
\begin{equation}
\label{eq:OmegaDM Starobinsky}
\Omega_{\chi} h^2 \simeq 8.9 \times 10^7 e^{-0.55 \pi \nu} \left(\frac{m_{\chi}}{\rm{GeV}} \right) \left(\frac{\Hend T_{\rh}}{\MP^2} \right) \quad \text{(Starobinsky inflation, $\alpha=1$)},
\end{equation}
which can be re-expressed in terms of the reheating temperature $T_\rh$ as
\begin{equation}
\label{eq:DM TRH Starobinsky}
T_{\rh} \simeq 
  8.0 \cdot 10^{12}\GeV \cdot \frac{10^{-9}}{e^{-0.55\pi \nu}} \cdot \frac{(10^{12}\GeV)^2}{m_{\chi} \Hend} 
  \quad \text{(Starobinsky inflation, $\alpha=1$)} \, .
\end{equation}%
\begin{figure}[h!]\centering%
Dark matter parameter space for Starobinsky inflation \vspace{-1em}\\
\includegraphics[width=0.49\textwidth]{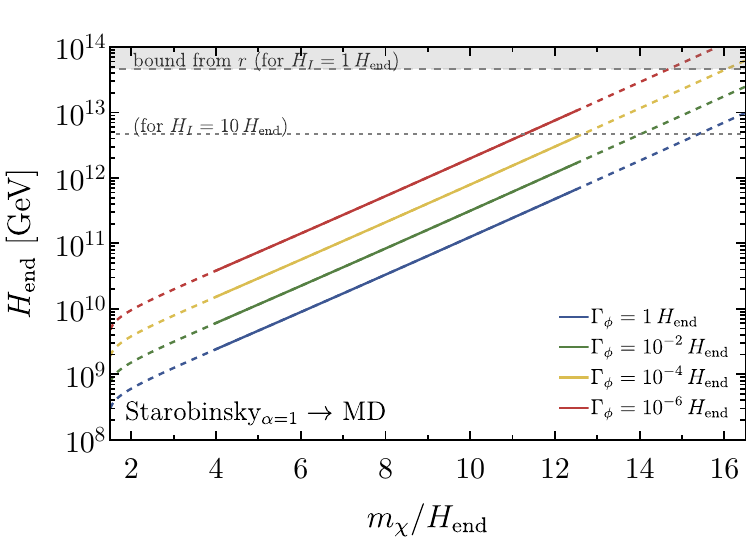} \hspace{-1em \hfill}
\includegraphics[width=0.49\textwidth]{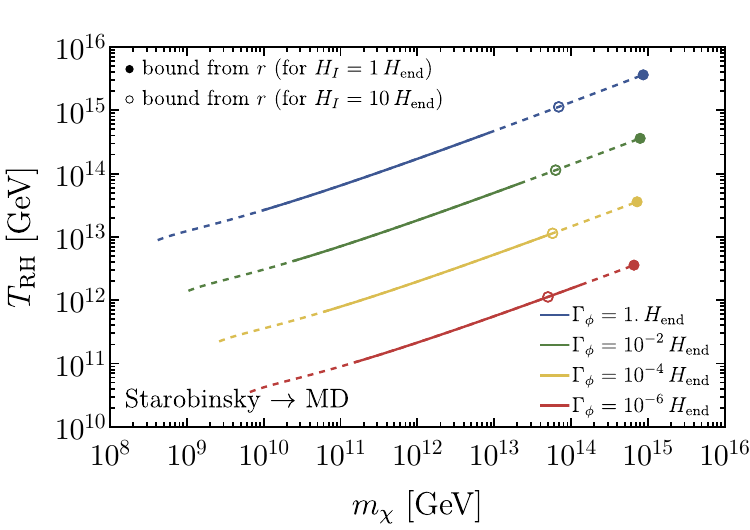}
\caption{Same as Fig.~\ref{fig:DM_dS}, for the model of Starobinsky inflation with $\alpha=1$, corresponding to heavy inflaton mass and predominant post-inflationary production.
The lines for the correct relic abundance of dark matter are shown with a solid line when they fall within the range of the points that we computed explicitly (shown in Fig.~\ref{fig:ntot_Starobinsky}), and with a dashed line when they rely on the extrapolation of \cref{eq:nco Starobinsky}.
}
\label{fig:DM_Starobinsky}
\end{figure}%
In conclusion, in a realistic inflationary model where post-inflationary production of a heavy spectator field could be larger than the inflationary production at Hubble crossing, the right mass range for DM increases from $\sim 1-4\,\Hend$ (\cref{fig:DM_dS})
to $\sim 2-14\,\Hend$ (\cref{fig:DM_Starobinsky}).

\section{Discussion and Conclusions}
\label{sec:conclusions}
We have investigated the gravitational production of superheavy scalar fields with mass $m_{\chi} \gtrsim \HI$ during and after inflation. 
This phenomenon arises inevitably due to gravitational interactions, and provides a minimal setup for dark matter production. 

We focus on an analytical approach to the calculation of the produced particle abundance, by means of the steepest-descent approximation. 
Besides the computational advantage in performing this calculation with respect to the numerical time evolution of the mode functions, we have highlighted the conceptual clarifications that emerge in this picture.
As a first step, the epochs of gravitational production can be easily identified by solving for $\omega_k(t)=0$ with complex $t$ (see \cref{fig:omega_zeroes}), pointing to the times around Hubble exit and at the end of inflation, when the inflaton oscillates in this minimum. In this approach, these different production phases emerge from a simple and coherent picture.
Second, the exponent in the suppressed particle abundance can be estimated directly from the expression of $\omega_k(t)$ (see \cref{fig:poles}). 
This allows to derive in a few lines known results for the particle occupation number ($e^{-2 \pi \nu}$ for de Sitter, with $\nu=\sqrt{m_\chi^2/H^2-9/4}$), and to derive novel results as \cref{eq:2pi nu(tk),eq:nk_reheating,eq:UV tail suppression} in a simple fashion.
We find that the production during inflation (with a slowly-evolving $H$) leads to $e^{-2 \pi \nu(t_k)}$ with $\nu(t_k)$ evaluated when the physical mode is still inside the Hubble radius (\cref{eq:2pi nu(tk)}), and that post-inflationary production can be captured by $e^{-\mathcal O(1) m_\chi/m_\phi}$ (\cref{eq:UV tail suppression}). As the inflaton mass $m_\phi$ increases, production after inflation becomes the dominant contribution.

We explored three inflationary scenarios: de Sitter transitioning to Minkowski, power-law inflation, and the Starobinsky model of inflation. 
The agreement between our analytical and numerical results validates this method and extends the applicability of our analytical techniques for various models of inflation. 
We notice that our results can be applied to the gravitational production of superheavy fermions and vectors, which obey a similar dispersion relation as scalars, leading to an exponentially-suppressed particle number density for inflationary production.

We also explored scenarios in which gravitationally produced heavy scalars are a viable dark matter candidate, assuming their stability over cosmological timescales. For superheavy scalar production with $m_{\chi} \gtrsim \HI$, the gravitational particle production is sensitive to the transition from the inflationary phase to the matter-dominated phase. We modeled the dynamics using the Starobinsky model of inflation, characterized by a variable parameter $\alpha$, that governs the steepness of this transition. 
For large-scale inflation ($\HI\gtrsim 10^9\GeV$) and relatively high reheating temperature, heavy spectators fields ($m_\chi\sim \mathcal O(1-10) \Hend$) can naturally account for the observed dark matter. 
Such a candidate is particularly compelling from a phenomenological perspective as they avoid isocurvature constraints, with $m_{\chi} \gtrsim 0.5 \HI$, and may lead to cosmological collider signatures or primordial gravitational waves (tensor modes), a prospect that future experiments will be able to explore.
Although the prospects for terrestrial detection of such a dark matter candidate are weaker, there are ideas in this direction as well.
The Windchime Project~\cite{Windchime:2022whs, Carney:2019pza} proposes to employ an array of mechanical sensors configured to function as a single detector. 
This setup is designed to detect subtle disturbances caused by superheavy dark matter as it interacts with these sensors. 
This experiment is most sensitive to masses approaching or exceeding the Planck scale, but it may be improved in the future to detect lower mass, possibly related to inflationary particle production.
Looking ahead, we eagerly anticipate developments in our theoretical understanding of inflationary physics and upcoming CMB and large-scale structure surveys, which can advance our understanding of the early universe, and provide us with new insights about the nature of dark matter.

\acknowledgments
We thank Valerie Domcke, Marcos A.~G.~Garc{\'i}a, Yohei Ema, Rocky Kolb, Marco Peloso, Riccardo Penco, Mathias Pierre, Michele Redi, Antonio Riotto, Leonardo Senatore and Richard Woodard for useful discussions and comments. D.R.\ is supported in part by NSF Grant PHY-2014215, DOE HEP QuantISED award \#100495, and the Gordon and Betty Moore Foundation Grant GBMF7946.
D.R.\ acknowledges hospitality from the Perimeter Institute for Theoretical Physics during the preparation of this paper. Research at Perimeter Institute is supported in part by the Government of Canada through the Department of Innovation, Science and Economic Development Canada and by the Province of Ontario through the Ministry of Colleges and Universities. S.V. and W.X. are supported in part by the U.S. Department of Energy under grant DE-SC0022148 at the University of Florida.

\appendix
\section{WKB Approximation}
\label{app:wkbapproximation}
In this appendix, we briefly discuss the higher-order WKB approximation. To ensure numerical precision in our particle abundance spectra, we primarily utilize the first improved WKB order in most of our plots and analyses. For a detailed discussion on generating higher-order terms in the adiabatic expansion, see Ref.~\cite{Dabrowski:2014ica}.

One can demonstrate that for an adiabatic expansion of the form
\begin{equation}
    \chi_k(t) \; = \; \frac{1}{\sqrt{W_k(t)}} e^{-i \int^t W_k} \, ,
\end{equation}
to satisfy the equation of motion~(\ref{eq:modeeq1}), it is necessary for $W_k(t)$ to satisfy the following constraint
\begin{equation}
    W_k^2(t) \; = \; \omega_k^2(t) + \sqrt{W_k(t)} \frac{d^2}{dt^2} \left(\frac{1}{\sqrt{W_k(t)}} \right) \, .
\end{equation}
Therefore, we can iteratively solve this expression to determine the $j$-th order of the adiabatic expansion. Up to the second improved order, we obtain
\begin{align}
    \omega^{(0)}_k(t) &=  \omega_k(t) \, ,  \\
     \omega_k^{(1)}(t) & =\omega_k(t)-\frac{1}{4}\left(\frac{\ddot{\omega}_k}{\omega_k^2}-\frac{3}{2} \frac{\dot{\omega}_k^2}{\omega_k^3}\right) \, ,  \\
     \omega_k^{(2)}(t) & = \omega_k(t)-\frac{1}{4}\left(\frac{\ddot{\omega}_k}{\omega_k^2}-\frac{3}{2} \frac{\dot{\omega}_k^2}{\omega_k^3}\right)-\frac{1}{8}\left(\frac{13}{4} \frac{\ddot{\omega}_k^2}{\omega_k^5}-\frac{99}{4} \frac{\dot{\omega}_k^2 \ddot{\omega}_k}{\omega_k^6}+5 \frac{\dot{\omega}_k \dddot{\omega}_k}{\omega_k^5}+\frac{1}{2} \frac{{\ddddot\omega}_k}{\omega_k} -\frac{297}{16} \frac{\dot{\omega}_k^4}{\omega_k^7}\right) \, .
\end{align}
At the $(j+1)$-th order of expansion, we find
\begin{equation}
\omega_k^{(j+1)} \; = \; \sqrt{\omega_k^2-\left[\frac{\ddot{\omega}_k^{(j)}}{2 \omega_k^{(j)}}-\frac{3}{4}\left(\frac{\dot{\omega}_k^{(j)}}{\omega_k^{(j)}}\right)^2\right]} \, ,
\end{equation}
truncated at the order of at most $2j$ derivatives with respect to time. Importantly, the final results evaluated at asymptotic time limit $t_f \rightarrow \infty$ are independent of the order of the WKB approximation, as the basis-dependent terms vanish in this limit.

\section{Stokes Line Approach}
\label{app:stokes}
In this appendix, we compare method of steepest descent approximation with the Stokes line approach. The latter was considered in the context of the Schwinger effect~\cite{Dabrowski:2014ica, Dabrowski:2016tsx}, and has been applied in the context of gravitational particle production~\cite{Li:2019ves, Li:2020xwr, Corba:2022ugu}. We note that when $m_{\chi} \gtrsim H_I$, from \cref{eq:omega_k} we see that $\nu^2 \gg 0$ and $\omega_k^2 \gtrsim 0$ is always positive. This allows us to approximate the location of the saddle points as
\begin{equation}
    \label{eq:scalefactorsaddle}
    a(t_0) \; \simeq \; \pm \frac{ik}{m_{\chi}} \, .
\end{equation}
In this work, we explored the WKB approximation for estimating the Bogoliubov coefficients and particle abundance~\cite{Kofman:1997yn, Chung:1998bt, Quintin:2014oea, Ema:2018ucl, Kaneta:2022gug} and focused on the steepest descent approximation. However, as argued in~\cite{Dumlu:2010ua}, the mode functions evolve substantially during the expansion of the universe and this necessitates to identify the dominant and subdominant contributions of the WKB approximation linked to particle production. As discussed in~\cite{Dabrowski:2014ica}, to fully account for the contribution arising from the negative frequency subdominant component, one needs to consider Stokes phenomenon. 

Recently, these concerns were studied in detail in Ref.~\cite{Corba:2022ugu}. The discrepancy between the WKB approximation and the Stokes phenomenon arises because the WKB adiabatic expansion can only be defined locally. This implies that mode functions at some initial time $t_i$ cannot be analytically extended across the entire complex plane, and the WKB approximation breaks down close to the poles (saddle points), where the adiabaticity conditions become violated. Thus, as the contour passes through a Stokes line, which characterizes the region where the local approximation is valid, the negative frequency contribution must be incorporated into the WKB solution.

The Stokes lines are determined by the condition
\begin{equation}
    F_k(t) \; \equiv \; -2i \int_{t_n}^t \omega_k(t') dt' \; = \; \rm{purely~real} \, ,
\end{equation}
where $t_n$ denotes the location of the saddle point $\omega(t_n) = 0$ lying in the lower half-plane. One can show that the Bogoliubov coefficient can be approximated as~\cite{Berry1989, Berry1990, Dabrowski:2014ica}
\begin{equation}
    \beta_k(t) \; \simeq \; -i S_k(t) e^{i \int_{t_n}^{t_n^*} \omega_k(t') dt'} \, ,
\end{equation}
with the Stokes multiplier defined as
\begin{equation}
    S_k(t) \; \equiv \; \frac{1}{2} \left[1 + \erf \left(-\frac{\Im F_k(t)}{\sqrt{2 |\Re F_k(t)|}} \right) \right] \, .
\end{equation}
Given that the error function varies between $-1$ and $1$, we observe that $\beta_k(t)$ transitions from $0$ to $i \, e^{i \int_{t_n}^{t_n^*} \omega_k(t') dt'}$. We now compare it with the steepest descent approximation. From Eq.~(\ref{eq:scalefactorsaddle}), we observe that for superheavy dark matter, the saddle points lie in the complex plane. Consequently, the exponential term can be estimated as $ e^{i \int_{t_n}^{t_n^*} \omega_k(t') dt'} =  e^{-2i \int_{\Re t_n}^{t_n} \omega_k(t') dt'}$. The integral in the exponential term coincides with Eq.~(\ref{eq:approxwkbint1}) employed in the steepest descent method. The primary difference between these methods lies in the prefactor. However, as the steepest descent approximation yields a prefactor of $\pi^2/9 \simeq 1$ for the coefficient $|\beta_k|^2$, it nearly matches the peak value of the Stokes multiplier $S_k(t) \simeq 1$, and the two approaches lead to nearly identical results. Consequently, in our analysis, we favor the simpler steepest descent approximation over evaluating the Stokes multiplier. For a comparison of the steepest descent method with the Stokes phenomenon, see Ref.~\cite{Enomoto:2020xlf}, and for a detailed exploration of superheavy scalar field production using the Stokes phenomenon, see Ref.~\cite{Li:2019ves}.

\bibliographystyle{JHEP}
\bibliography{bib_GPP-Scalar}

\end{document}